%% file: CaIIDepletionArxiV_GSardane.tex
\documentclass[a4paper,13pt]{mn2e} 
\pdfoutput=1
\usepackage{graphicx}
\usepackage{deluxetable} 
\usepackage{lscape} 
\usepackage{amssymb,amsmath}
\usepackage{cite}
\usepackage{csquotes}
\usepackage{color,soul}
\usepackage{multirow}
\usepackage{booktabs}
\usepackage{threeparttable}
\usepackage{wasysym}
\usepackage[utf8]{inputenc}

\newcommand{\HI}{\mbox{H\,{\sc i}}} \newcommand{\MgII}{\mbox{Mg\,{\sc ii}}}
\newcommand{\MgI}{\mbox{Mg\,{\sc i}}} \newcommand{\FeII}{\mbox{Fe\,{\sc ii}}}
\newcommand{\CIV}{\mbox{C\,{\sc iv}}} 
\newcommand{\CrII}{\mbox{Cr\,{\sc ii}}}
\newcommand{\ZnII}{\mbox{Zn\,{\sc ii}}}
\newcommand{\CoII}{\mbox{Co\,{\sc ii}}}
\newcommand{\CdII}{\mbox{Cd\,{\sc ii}}}
\newcommand{\FeI}{\mbox{Fe\,{\sc i}}}

 \newcommand{\MnII}{\mbox{Mn\,{\sc ii}}}
 \newcommand{\AlIII}{\mbox{Al\,{\sc iii}}}

\newcommand{\NiII}{\mbox{Ni\,{\sc ii}}}

\newcommand{\SiI}{\mbox{Si\,{\sc i}}} 
\newcommand{\SiII}{\mbox{Si\,{\sc ii}}}

\newcommand{\CaI}{\mbox{Ca\,{\sc i}}}
\newcommand{\CaII}{\mbox{Ca\,{\sc ii}}}
\newcommand{\TiII}{\mbox{Ti\,{\sc ii}}}
\newcommand{\NaI}{\mbox{Na\,{\sc i}}}                                                        

\input{jour}
\title[\CaII~Absorbers in the
SDSS]{\CaII~Absorbers in the Sloan Digital Sky Survey: Element Abundances and Dust} 
\author[G. M.Sardane et al.] {\parbox[t]{\textwidth}{\raggedright Gendith M. Sardane$^{1}$\thanks{E-mail:
gms48@pitt.edu}, David A. Turnshek$^{1}$, and Sandhya M. Rao$^{1}$} \vspace*{6pt}\\
$^{1}$Department of Physics and Astronomy and PITTsburgh Particle physics, Astrophysics, and
Cosmology Center (PITT PACC),\\ University of Pittsburgh, Pittsburgh, PA 15260\\ }

\begin{document}

\date{}

\pagerange{\pageref{firstpage}--\pageref{lastpage}} \pubyear{2014}
\maketitle

\label{firstpage}

\begin{abstract} We present measurements of element abundance ratios and dust in \CaII~absorbers
identified in SDSS DR7+DR9.  In an earlier paper we formed a statistical sample of 435 \CaII\ absorbers and postulated that their statistical properties might be representative of at least two  populations of absorbers. Here we show that if the absorbers are roughly divided into two subsamples with \CaII\ rest equivalent widths larger and smaller than $W_0^{\lambda 3934} = 0.7$ \AA, they are then representative of two physically different populations. Comparisons of abundance ratios between the two \CaII\ absorber populations indicate that the weaker $W_0^{\lambda 3934}$ absorbers have properties consistent with halo-type gas, while the stronger absorbers have properties intermediate between halo- and disk-type gas. We also show that, on average, the dust extinction properties of the overall sample is consistent with a LMC or SMC dust law, and the stronger absorbers are nearly 6 times more reddened than their weaker counterparts.  The absorbed-to-unabsorbed composite flux ratio at $\lambda_{rest} = 2200$ \AA\ is $\mathcal{R} \approx 0.73$ and $E(B-V) \approx 0.046$ for the stronger \CaII\ absorbers ($W_0^{\lambda 3934} \ge 0.7$ \AA), and $\mathcal{R} \approx 0.95$ and $E(B-V) \approx 0.011$ for the weaker \CaII\ absorbers ($W_0^{\lambda 3934} < 0.7$ \AA).
\end{abstract}


\section{Introduction} \label{intro} 
Determination of element abundances, dust properties, and the overall chemical histories of the gaseous environments of galaxies is needed for an improved understanding of galaxy formation and evolution. The gaseous environments of galaxies include their interstellar medium (ISM) as well as surrounding circumgalactic medium and intergalactic medium (CGM and IGM). A broad goal is to constrain how galaxies convert their gas into stars, and the feedback (outflow) mechanisms that are at play in polluting the gas surrounding galaxies in the context of the observed galaxy stellar populations. Here we consider what can be learned from a study of the properties of intervening gas giving rise to \CaII\ absorption seen in the spectra of background quasars. The statistics of these absorbers, as derived from an analysis of SDSS quasar spectroscopy, were recently presented in Sardane, Turnshek \& Rao (2014; hereafter Paper I). Some of the properties of galaxies associated with them, derived from SDSS images, will be presented in Sardane, Turnshek \& Rao (2015; hereafter Paper III). In this contribution (Paper II in the series), we use SDSS spectroscopy to constrain results on relative element abundances and dust in \CaII\ absorbers. 

Quasar absorption line (QAL) spectroscopy is a unique probe of the evolution of galaxies 
and their gaseous components, from the coolest molecular clouds to the hotter ionized gaseous halos (Foltz et al. 1988; Lu et al. 1996; Ledoux et al. 1998; Wolfe \& Prochaska 2000a; Petitjean et al. 2000; Ledoux, Srianand \& Petitjean  2002; Ledoux, Petitjean, \& Srianand 2002; Cui et al. 2005; Srianand et al. 2005; Petitjean et al. 2006; Fox et al. 2007; Tripp et al. 2008; Tom \& Chen 2008; Noterdaeme et al. 2008; 2010; Prochaska et al. 2011; Tumlinson et al. 2011a; Crighton et al. 2013; Stocke et al. 2014; Lehner et al. 2014; Savage et al. 2014).
Due to the rest-frame UV location of the resonance transitions most relevant for QAL studies, such studies 
have traditionally concentrated on
probing the gaseous absorbers and their environments at high redshifts. As a result, the paucity of identified gaseous 
structures at very low redshift naturally creates a gap in our understanding of how galaxies and their gaseous 
environments evolve from high redshift to the present. Moreover, cosmological dimming, which reduces the surface brightness
of astronomical sources by $(1+z)^4$, makes it more difficult to
identify and characterize the galaxies that could host the absorbers at high redshift.

%%Here I wanted to talk about the usefulness of the CaII transition
One rare class of QAL system that is not as well studied and understood as others is the one identified using the resonance 
doublet transition of singly ionized calcium: \CaII~$\lambda\lambda 3934, 3969$. 
However, although its incidence makes it rare, the advantage of studying the \CaII~QAL doublet is that it can be observed from $z\sim 1.4$ all the way down to the present epoch using the large number of optical ground-based quasar spectra obtained by the Sloan Digital Sky Survey (SDSS) (Schneider et al. 2010; Ahn et al. 2012).

Recently, we harnessed the statistical power of the 
SDSS to assemble the largest catalog of these rare \CaII~absorbers (Paper I). This search, which utilized $\sim 95,000$ 
quasar sightlines from the Seventh (DR7) and Ninth (DR9) data releases of the SDSS, resulted in the
compilation of 435 \CaII~absorbers. As described in Paper I, the detections
were based on $\geq5$$\sigma$ and $\geq2.5$$\sigma$ rest equivalent width significance threshholds for the 
strong and weak members of the \CaII~doublet, respectively. A constraint on the doublet ratio was also employed
to remove \enquote{unphysical} profiles, as dictated by the theoretical ratio of the doublet 
oscillator strengths.

In Paper I we demonstrated that after accounting for sensitivity corrections, 
a single power-law fit is insufficient to describe the $\lambda \mathrm{3934}$ rest equivalent width, 
$W_0^{\lambda3934}$, distribution. More specifically, a two-component exponential 
distribution is required to fit the data satisfactorily.
This result is somewhat surprising based on analysis of much larger samples of more ubiquitous
QAL systems such as \MgII~(e.g., Nestor, Turnshek \& Rao et al. 2005, Seyffert et al. 2013, and 
Zhu \& Menard 2013) and \CIV~(Cooksey et al. 2013).
For these QAL systems, a single exponential function suffices
to characterize their $\mathrm{W_0}$ distributions at $\mathrm{W_0~\gtrsim 0.1~\AA}$. 
For \CaII~absorbers the need for a two-component distribution
persists across all observed redshifts, which is strong statistical 
evidence for \textit{at least} two distinct populations. 

A preliminary investigation of the nature of the two-component fit suggested that there was a bimodality in the distribution of
\MgII-to-\CaII~ratios (i.e., $\mathrm{W_0^{\lambda2796}/{W_0^{\lambda3934}}}$) with a separation above and below 
$W_0^{\lambda3934} \sim 0.7$ \AA. However, there was no evidence that the \CaII\ doublet ratio, 
which is an indicator of saturation, could be used to distinguish between the two absorber populations.

Using the statistical sample from Paper I, this work
will explore the chemical abundances and dust-extinction properties of the \CaII\ absorbers.
In particular, we will exploit the power of spectral stacking to form various composite spectra which will be analyzed to infer chemical 
abundance ratios and dust-extinction properties for various subsamples. This will allow us to characterize and distinguish between 
two different \CaII\ absorber populations as implied by their statistical properties.

The paper is organized as follows. In \S2 we give a brief description of our SDSS \CaII\ absorber catalog that was presented in Paper I. 
In \S3 we discuss notable individual systems in the \CaII\ catalog. In \S4 we derive the composite
properties of the \CaII\ absorber full sample and subsamples in the context of their element abundance ratios and dust-extinction properties. We then discuss the 
implications of these results and how they explain the existence of two different populations of \CaII\ absorbers in \S5.
In \S6 we summarize our results and conclusions.
 
\section{The SDSS \CaII~Absorber Catalog} \label{data} 
The sample of \CaII~absorbers used in this analysis is derived from our Paper I catalog. 
It consists of 435 \CaII~absorbers with $W_0^{\lambda3934} \geq 0.16 \mathrm{~\AA}$, compiled using over 95,000 quasar spectra with SDSS magnitudes $i < 20$ from the SDSS data releases DR7 
(Abazajian et al. 2009; Schneider et al. 2010) and DR9 (Ahn et al. 2012; P\^aris et al. 2012). Data from DR7 and DR9 were obtained using two nearly identical spectrographs, the SDSS spectrograph and the Baryon Oscillation Spectroscopic Survey (BOSS) spectrograph, respectively. 
The BOSS spectrograph (Smee et al. 2013), which was designed to target higher-redshift quasars 
for the  BOSS project (Schlegel et al. 2007; Dawson et al. 2013),
is an improved version of the SDSS spectrograph. The SDSS spectrograph 
covers the wavelength range of $\mathrm{3800 - 9200~\AA}$, while the BOSS spectrograph 
has extended wavelength coverage in both the blue
and the near-infrared, and covers $\mathrm{3600 - 10,400~\AA}$. The resolutions of both spectrographs are essentially the same, 
ranging from $\sim1500$ at $3800~\mathrm{\AA}$ to $\sim2500$ at $9000~\mathrm{\AA}$. 
%The search for \CaII\ absorption, however, is limited to $3800~\mathrm{\AA}$ through $9200~\mathrm{\AA}$.

To identify the \CaII\ absorbers, splines and Gaussians were used to fit a quasar spectrum's so-called pseudo-continuum, which consists of  the \enquote{true} continuum plus the broad emission lines.
As indicated previously, the absorbers were selected based on $5\sigma$ and $2.5\sigma$ 
significance thresholds for the $\lambda3934$~and $\lambda3969$~lines, respectively, and a 
doublet ratio (DR)\footnote{The doublet ratio is defined here as 
$\mathrm{DR = W_0^{\lambda3934}/W_0^{\lambda3969}}$.} constraint of $\mathrm{1 \leq DR \leq 2}$ to within the measurement errors, which is the range of physically-allowable doublet ratios
between saturated (DR = 1) and completely unsaturated (DR = 2) absorption lines.

\section{Properties of Some Individual \CaII~Absorbers in the Catalog}

In the limit of an absorption line in the optically thin regime, the optical depth is independent of the 
Doppler parameter, so a measurement of its equivalent width translates reliably into a column density measurement.
Hence, for weak, unsaturated resonance transitions at rest-frame wavelength, $\lambda_0$, and oscillator strength, $f$, 
the column density, $N$, is approximately a linear function of the rest equivalent width, $W_0$, 

\begin{equation} 
N \approx 1.13 \times 10^{20}~\frac{W_0}{\lambda_0^{2} f}
\label{linearCOG}
\end{equation}

\noindent
where $N$ is in atoms cm$^{-2}$ and $W_0$ and $\lambda_0$ are in \AA\ (Draine 2011). 

In QAL studies of high-N(\HI) systems, which generally applies to the \CaII\ absorbers, it is common to use this relation to derive total element column densities from weak, low-ionization, unsaturated lines due to, e.g., \ZnII, \CrII, \FeII, and \MnII. Self-shielding generally ensures that the low-ionization metallic elements will be the dominant ionization state. Therefore, we will employ this method to infer some of the properties of individual \CaII\ absorbers, and we will also take advantage of this in the \S4 analysis using composite spectra. These assumptions are theoretically justified, and departures should be small for the elements we consider (e.g., Viegas 1995; Vladilo et al. 2001; Prochaska \& Wolfe 2002). In cases where there is evidence for a non-negligible degree of saturation, we will report lower limits on derived column densities.

Under these assumptions, we derive abundance ratios of special sightlines that have high enough redshift and signal-to-noise ratios to permit spectral coverage and reliable measurements of interesting weak absorption features. In keeping with standard practice, the reported abundance ratios are relative to solar (Asplund et al. 2009), i.e., the abundance ratio of element X relative to element Y will be given relative to solar values:

\begin{equation} 
[X/Y] \equiv \log[N(X)/N(Y)] - \log[N(X)/N(Y)]_{\astrosun} 
\end{equation}

For absorbers with $z_{abs} \gtrsim 0.9$, the UV \ZnII-\CrII~rest-frame region of the spectrum falls into the SDSS optical wavelength window. The \CaII~sample consists of $\sim 70$ systems with $z_{abs} \gtrsim 0.9$. However, due to increasingly poor signal-to-noise ratios in the blue region of many SDSS spectra, and the unfortunate blending of the \ZnII-\CrII~region with Ly$\alpha$ forest lines or other unrelated metal lines, the useful sample where \ZnII-\CrII\ can be studied in \CaII\ absorbers is reduced to a dozen systems. Since zinc is only mildly refractory, its abundance ratio relative to more strongly depleted elements such as chromium, titanium, and iron is of primary importance for characterizing the depletion properties of the gas.

The \ZnII-\CrII~region has a rest-frame wavelength interval $\mathrm{2026-2066~\AA}$. For the usable spectra we infer the column densities of Cr and Zn from four \CrII~ transitions and two \ZnII~transitions. The first of these is a feature at $\lambda$2026 which is a blend due to three transitions: a \ZnII\ line, a weak \MgI\ line, and a very weak \CrII\ line. Another feature at $\lambda$2062 is a blend of \CrII\ and \ZnII. The two additional transitions for \CrII~are at $\lambda2056$ and $\lambda2066$.

Deblending the 
features in the \ZnII-\CrII\ region at SDSS resolution can be done by making use of known oscillator strength ratios for an element's ionic transition. For example, for the feature at $\lambda2026$ the equivalent widths of \MgI\ $\lambda2026$ and \CrII\ $\lambda2026$ were taken to be 32 and 23 times smaller than the observed equivalent widths of the unblended \MgI\ $\lambda2852$ and \CrII\ $\lambda2056$ lines, resepctively. The remaining absorption can then be attributed to \ZnII\ $\lambda2026$.\footnote{In some systems comparisons of \MgI\ $\lambda2852$ to \MgII\ $\lambda\lambda2796,2803$ suggest that \MgI\ $\lambda2852$ may be approaching saturation, but neglecting this does not introduce a significant uncertainty.} Generally, since the oscillator strength of \CrII\ $\lambda2026$ is quite small relative to \ZnII\ $\lambda2026$, its contribution to the $\lambda2026$ feature is negligible.

Similarly, the feature at $\lambda2062$ due to \ZnII\ and \CrII\ can be deblended by taking the \CrII\ $\lambda2062$ equivalent width to be half of the sum of the \CrII\ $\lambda2056$ and \CrII\ $\lambda2066$ equivalent widths, with the remainder due to \ZnII\ $\lambda2062$. The corresponding errors are then propagated in quadrature. The reported column densities are the error-weighted average values inferred from each transition. The results on the column densities for Zn$^+$, Cr$^+$, Fe$^+$ and Mn$^+$ are summarized in Table \ref{TableFive}, where the column density for Fe$^+$ is inferred from the weak lines of \FeII\ $\lambda\lambda2249,2260$ and the column density for Mn$^+$ is inferred from the weak lines of \MnII\ $\lambda\lambda\lambda2576,2594,2606$.
%What does the \CaII/\NaI ratio tell us?

\begin{figure*} 
\includegraphics[width=0.75\textwidth]{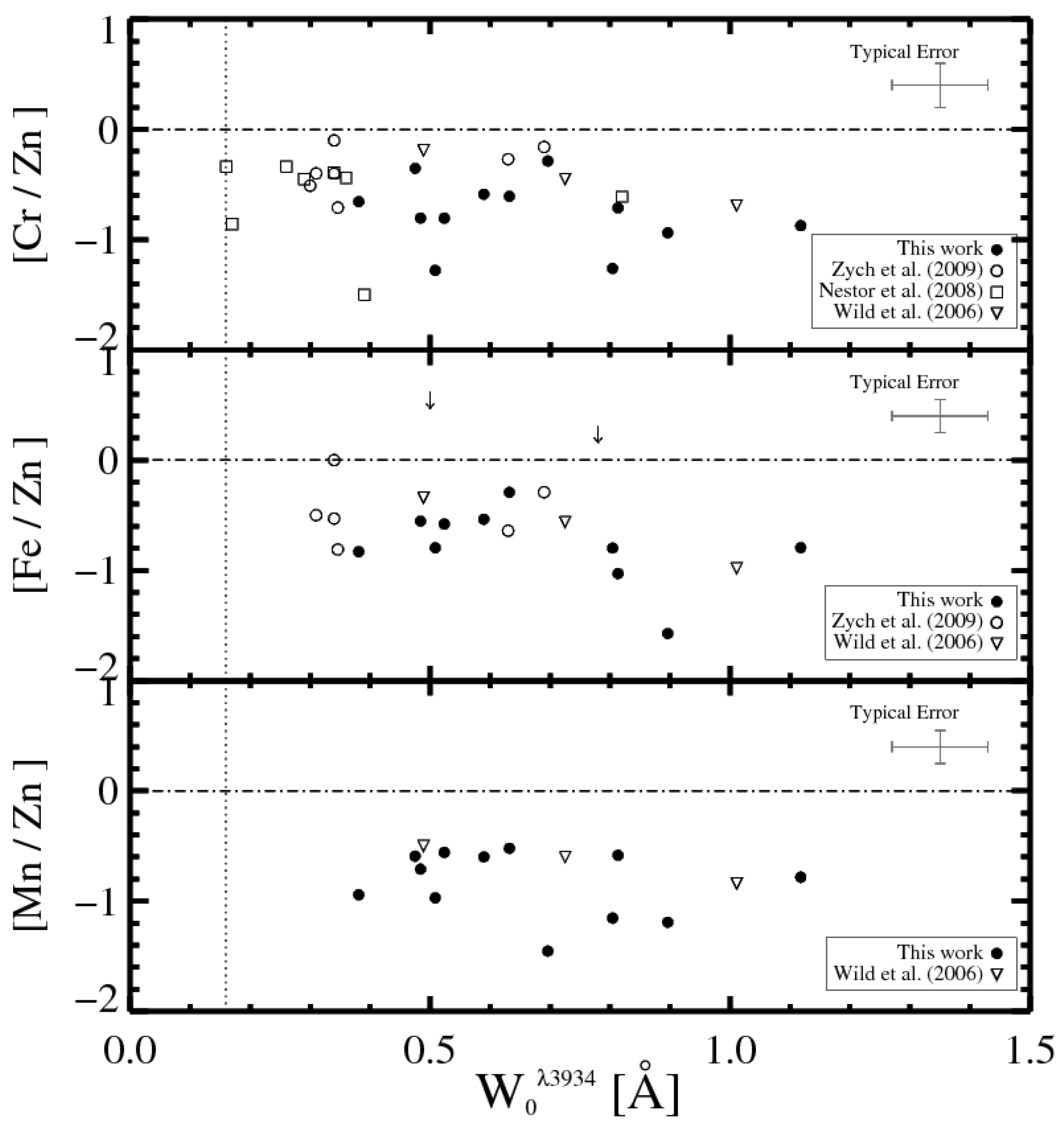}
\caption{
\textit{Top panel:} The abundance ratio [Cr/Zn] vs. $W_0^{\lambda3934}$ for the dozen individual \CaII\ absorbers in our sample that have reliable \CrII\ and \ZnII\ measurements (filled circles).  
\textit{Center panel:} The abundance ratio [Fe/Zn] vs. $W_0^{\lambda3934}$ (filled circles), as inferred from \FeII\ $\lambda\lambda2249,2260$. The downward arrows are upper limits from Zych et al. (2009). 
\textit{Bottom panel:} The abundance ratio [Mn/Zn] vs. $W_0^{\lambda3934}$ (filled circles) as inferred from \MnII\ $\lambda\lambda\lambda2576,2594,2606$. 
Typical errors are shown in the upper right of all three panels. Open symbols represent results from the other studies. Note that the Wild et al. (2006) data points are for their full sample, and high, and low  $W_0^{\lambda3934}$ composites. All three results show evidence for increasing depletion with increasing $W_0^{\lambda 3934}$, although the spread in [X/Zn] is large. The dotted vertical line in the figure marks the lowest values of the \CaII\ rest equivalent width detection threshold from Paper I, i.e.,  $W_0^{\lambda3934}$ = 0.16 \AA. The dash-dot horizontal lines mark the solar reference level for no depletion. The error bars on the right side of each panel indicate the typical uncertainties for these measurements. }
\label{PlotIndividual} 
\end{figure*}

For the \CaII\ absorbers in Table \ref{TableFive}, Figure \ref{PlotIndividual} shows the abundance ratios [X/Zn], where X = [Cr, Fe, Mn], as filled circles versus $W_0^{\lambda 3934}$. Cr, Fe, and Mn are known to be highly refractory, while Zn is not. Hence,
an indication of the degree of depletion of Cr, Fe, and Mn onto dust grains can be inferred from these abundance ratios. Results from previous investigations of \CaII\ absorbers (open symbols) known to be damped Ly$\alpha$ absorbers (DLAs) and subDLAs at $z\sim1$ are also included in Figure \ref{PlotIndividual} for comparison. These data are due to Nestor et al. (2003), Wild et al. (2006)\footnote{Wild et al. (2006) abundance ratios were derived from composite spectra constructed from 37 \CaII~absorbers from SDSS DR3. One of their results was for their entire sample, while two additional result were presented for \enquote{High} and \enquote{Low} $W_0^{\lambda 3934}$ values on either side of their median $W_0^{\lambda3934} = 0.68~\mathrm{\AA}$ value.}, and the Zych et al. (2009) VLT/UVES and Keck/HIRESb datasets.  Consistent with previous results, these refractory elements are all seen to be depleted relative to Zn. While the depletion levels are fairly typical, nucleosynthetic processes can yield departures of $-0.3$ to $-0.1$ in [Fe/Zn] (Prochaska \& Wolfe 2002). The depletion levels are seen to vary over a range of values, suggestive of a significant range in dust-to-gas ratios in these sightlines and/or environments.

Although a large scatter is present, the results suggest that depletion increases with increasing $W_0^{\lambda3934}$, which is generally consistent with previous findings.

\input{TableFive}

\input{TableSix}

The \NaI\ $\lambda\lambda$5891,5897 absorption transitions can be observed with the SDSS spectrograph for absorbers with $z_{abs} \lesssim 0.6$. With the BOSS spectrograph the coverage is extended to $z_{abs} \lesssim 0.7$. For absorbers in the \CaII\ catalog, measurements of \NaI\ are possible for 213 systems. However, due to signal-to-noise limitations, which is particularly severe for many of the \NaI\ lines since they occur close to the red limit of SDSS/BOSS spectra, only $31$ \CaII\ absorbers had $\geq 2\sigma$ \NaI\ detections. The results are summarized in Table \ref{TableSix}. However, based on the observed doublet ratios, 23 of the measurements indicate some degree of saturation, and so most of the results are given as lower limits on Na$^+$ column densities. For the remaining eight, Eq. \ref{linearCOG} was employed to derive the  Na$^+$ column densities. 

Galactic ISM studies show that Na$^+$ to Ca$^+$ column density ratios span about four orders of 
magnitude, ranging from $\sim +2.5$ to $\sim -1.5$ dex  (Routly \& Spitzer 1952; Siluk \& Silk 1974; 
Welty et al. 1996), which is mainly due to substantial differences in 
the depletion of Ca onto dust grains. Large ratios occur in cold, dense and quiescent clouds, whereas the smaller values can be attributed to environments where no significant depletion has yet occurred or where some Ca has been returned to the gas phase due to shocks, such as those in warm and/or high velocity clouds.

%The sample 
Finally, we note that our \CaII\ catalog has two cases where the \CaII\ and \NaI\ lines have 
doublet ratios which are  clearly indicative of lines in the unsaturated regime. Both of 
these have $W_0^{\lambda 3934} < 0.6 \mathrm{\AA}$, and in those cases we calculate the 
Na$^+$/Ca$^+$ column density ratios to be $-0.36 \pm 0.06$ dex and $-0.04 \pm 0.10$ dex. 
These values fall within the range that is typical of the diffuse, warm neutral medium in 
the Milky Way, where T $= 10^2 - 10^4$ K and n$_H \leq 10$ atoms cm$^{-3}$ (Crawford 1992; 
Welty et al. 1996; Richter et. al 2011). 
%For these physical conditions, Ca$^{+1}$ and Na$^{0}$ might not be the dominant ionic states. 

%%% % %% % % % % % %% % % % % % %% % % % % % % % % % % %% % % % % % %% % % % % % % % % % % % %

\section{Properties of \CaII~Absorbers from Composite Spectra}
Here we explore the composite properties of the \CaII\ absorbers by considering the 435 \CaII\ absorbers in the Paper I catalog. We form two types of composite spectra. The first type is one
constructed by median-combining continuum-normalized spectra. Element abundances will be inferred from this type of composite spectrum. The second type is formed using the geometric mean of unnormalized flux spectra (e.g., York et al. 2006, Vanden Berk et al. 2001). From this type of composite we will measure the overall extinction and reddening characteristics of the \CaII\ absorbers. This is done using unabsorbed quasar spectra that are matched (in emission redshift and $i$-band magnitude) to the \CaII\ absorber quasar spectra.  

We form these two types of composites for our full sample and four different subsamples, which are subsets of the full sample. 
The rationale behind choosing the criteria to define the four subsamples was given in Paper I. In particular, Paper I showed the existence of two populations of \CaII\ absorbers that could be separated based upon the bimodality in the distribution of   W$_0^{\lambda2796}$/W$_0^{\lambda3934}$ ratios, with the separation occurring at $\rm{W}_0^{\lambda 3934} = 0.7$\AA. We also found that the change in slope of the $\rm{W}_0^{\lambda 3934}$ distribution occurred at W$_0^{\lambda2796}$/W$_0^{\lambda3934} = 1.8$. (See figures 17 and 18 of Paper I.) Therefore, we define the four subsamples as systems with $\rm{W}_0^{\lambda 3934}$ $< 0.7$\AA\ and $\geq0.7$\AA, and those with W$_0^{\lambda2796}$/W$_0^{\lambda3934}$ $< 1.8$ and $\geq 1.8$. 
The results we derive from the full sample are primarily discussed in this section, while the results from the four subsamples are primarily discussed in \S5.

%We will discuss details of the latter in \S4.2.

\subsection{Normalized Composite Spectra}

The stacking procedure begins by shifting all 435 normalized spectra to the \CaII\ absorber rest frame. 
To facilitate wavelength registration before doing this, we start by rebinning the spectra into a finer sub-pixel grid, the size of which is about 
one-tenth of the original pixel size. Since the median is a robust measure of central 
tendency which properly gives higher weight to spectra with lower noise, we chose to median-combine the spectra to build the normalized composite (Vanden Berk et al. 2001; Pieri et al. 2010). 
% rather than average-combine.
The error in the composite flux is estimated  using the absolute deviations about the median flux (MAD), which is a robust estimator of variance, (e.g. Rousseeuw \& Croux 1993;  Lee-Brown et al. 2015). The final stack is then rebinned to 2-pixels per resolution element and then smoothed over two pixels for display purposes only.  

\begin{figure*} 
\includegraphics[width=0.95\textwidth]{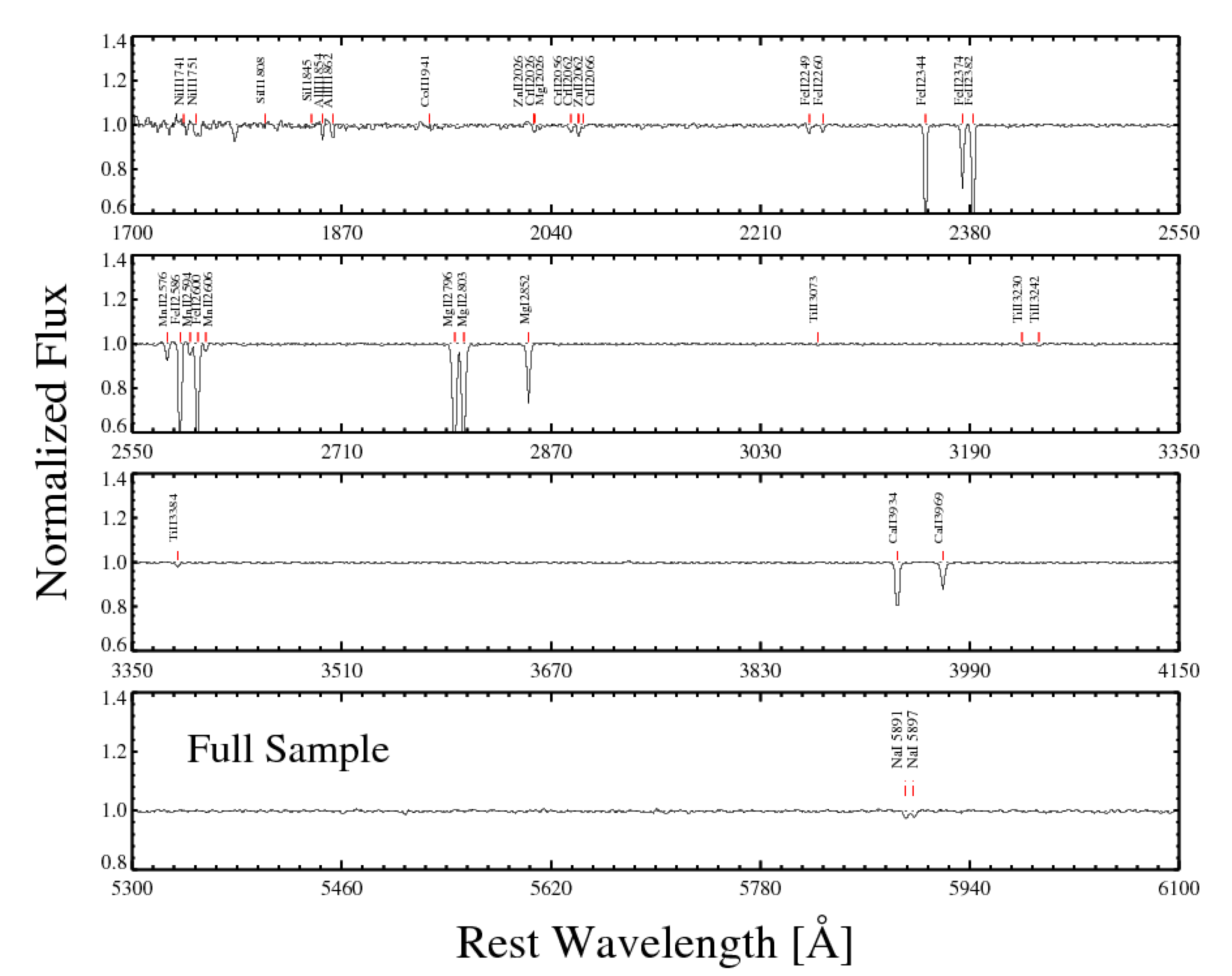} 

\caption{The median-combined normalized composite spectrum of the full sample of \CaII\ absorbers in the Paper I catalog. 
To facilitate accurate wavelength registration, each spectrum in the composite has been shifted to the rest frame of the \CaII~absorber prior 
to stacking using a finer subpixel grid. Details are provided in the text.
% that is a tenth of the size of the original pixel size. The final composite is then rebinned to two pixels per resolution element and then smoothed over two pixel. The error array is estimated by taking the median of the absolute deviations from the composite flux. Averaged over the entire wavelength range, the error in the flux is at $4\%$ of the flux. 
Red vertical lines mark absorption features that are significant at the $>2\sigma$ level. 
%A possible weak emission feature from a forbidden transition of [$\OIII$]$\lambda\lambda$372,3729, which would be an indication of star formation, is labeled using a blue vertical line. 
The rest equivalent width measurements and $2\sigma$ upper limits of QALs associated with the \CaII\ absorption in this spectrum are summarized in Table \ref{TableOne}.}
\label{NormComposite400} 
\end{figure*}

Figure \ref{NormComposite400} shows the full sample normalized composite spectrum, which was formed using all 435 SDSS \CaII\ absorber spectra
identified in Paper I. Figures $3-6$ are the normalized composite spectra for the four subsamples (see \S5 discussion). The error array of the full sample composite spectrum ranges between  $\sim$2\% and $\sim 8$\% of the flux. The red vertical lines mark the rest-frame locations of the absorption features, which are typical of those
 identified in QAL studies. The wavelength coverage of the various spectra 
in the \CaII\ absorber rest frame allows access to  absorption features that lie between and include \SiII\ $\lambda$1808 and \NaI\ $\lambda \lambda$5891,5897. Note that the stack includes absorber rest frame wavelengths down to $\sim 1700~\mathrm{\AA}$, but we do not attempt to measure any QALs at $\lambda \lesssim 1750$ \AA\  because of potential errors in the continuum placement at the (noisy) blue end of SDSS spectra. 
Clearly seen in the normalized composite spectrum are the transitions of low-ionization 
lines such as \ZnII, \CrII, \NiII, \TiII, \FeII, \MnII\, \MgII, \MgI\ and \NaI, as well as the higher-ionization transitions due to \AlIII. In the individual SDSS spectra most of these transitions are too weak, and/or located in spectral regions with too poor signal-to-noise, to identify them. In the \ZnII-\CrII\ region at shorter wavelengths, the composite is derived from only $\sim60$ out of $>400$~spectra, and thus it exhibits poorer signal-to-noise characteristics;
on the other hand, at the longer wavelengths the \NaI~region composite is comprised of $\sim200$ spectra. In Figure \ref{NormComposite400} we only label those features with $\geq 2\sigma$ detections to avoid crowded labeling. We measure the rest equivalent widths by fitting Gaussian profiles to each feature, with the constraint that features have a minimum line width set by the resolution of the SDSS/BOSS spectra. The rest equivalent widths of the various QAL transitions are summarized in Table \ref{TableOne}. For those lines which do not pass the 2$\sigma$ detection limit, we report $2\sigma$ upper limits. Composite results for the full sample and four subsamples are given in Table \ref{TableOne}. 
  
\begin{figure*} 
\includegraphics[width=0.95\textwidth]{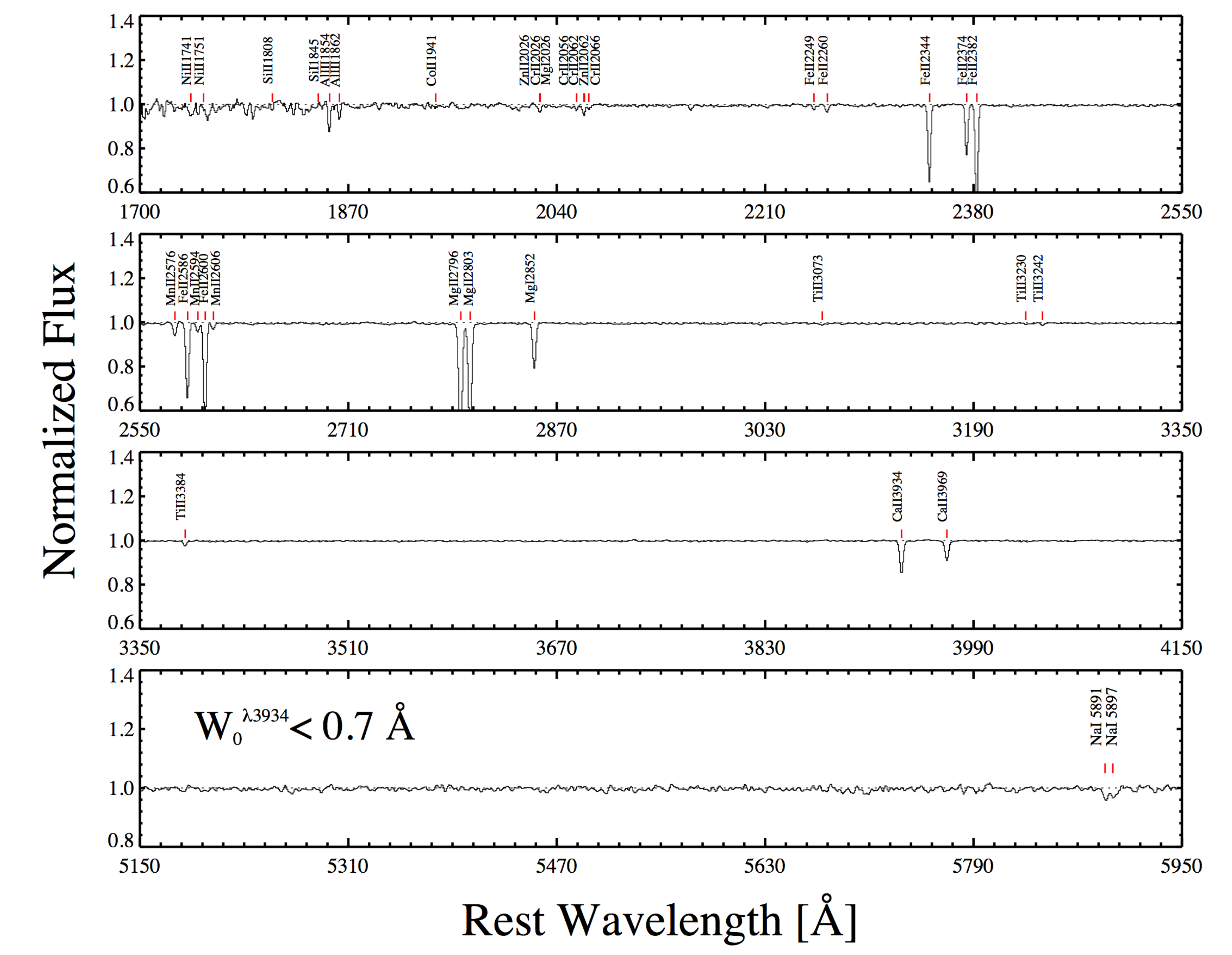} 
\caption{Same as Figure 2 but for the subsample of \CaII\ absorbers with W$_0^{\lambda3934} < 0.7$ \AA.}
\label{Fig3} 
\end{figure*}

\begin{figure*} 
\includegraphics[width=0.95\textwidth]{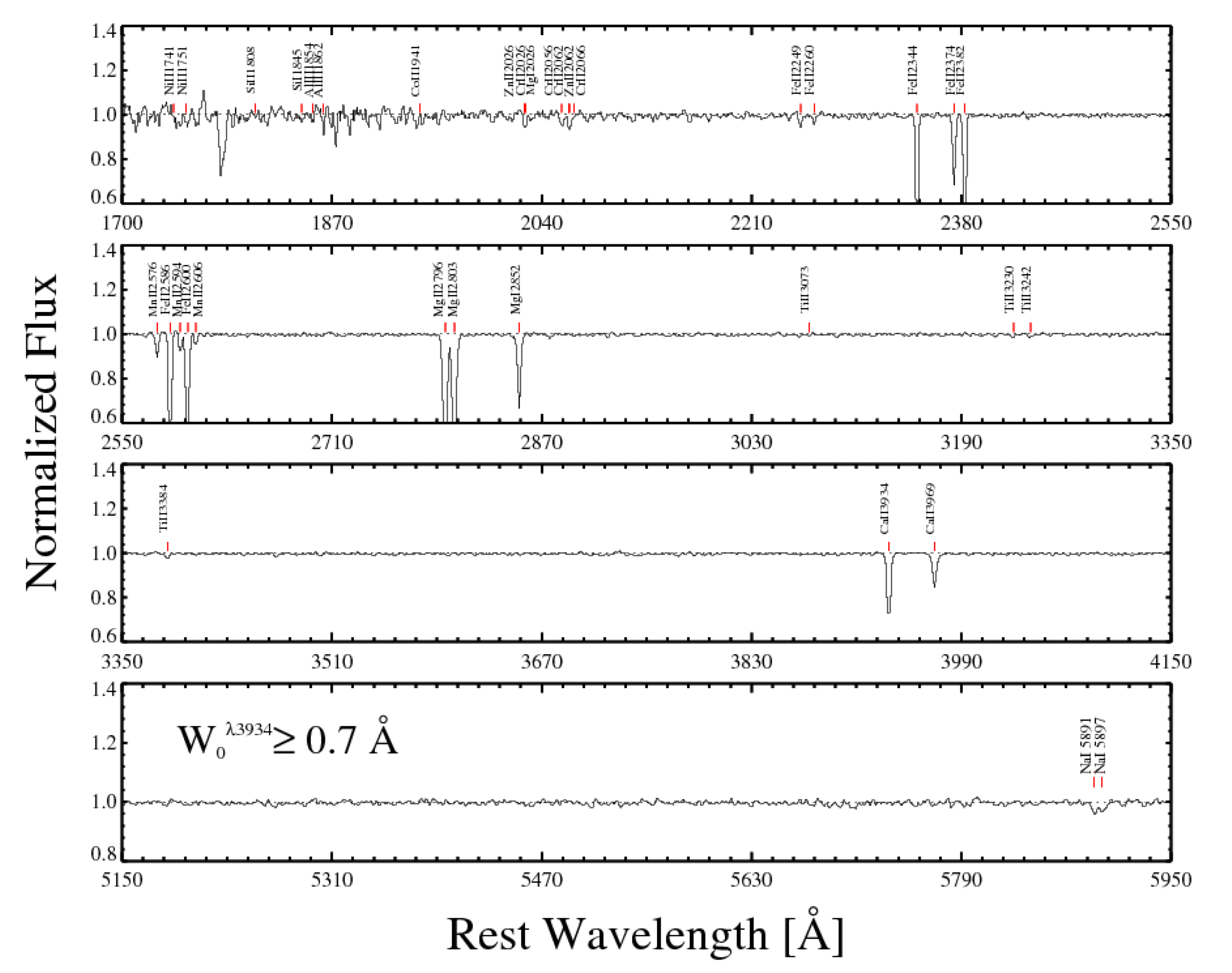} 
\caption{Same as Figure 2 but for the subsample of \CaII\ absorbers with W$_0^{\lambda3934} \ge 0.7$ \AA.}
\label{Fig4} 
\end{figure*}

\begin{figure*} 
\includegraphics[width=0.95\textwidth]{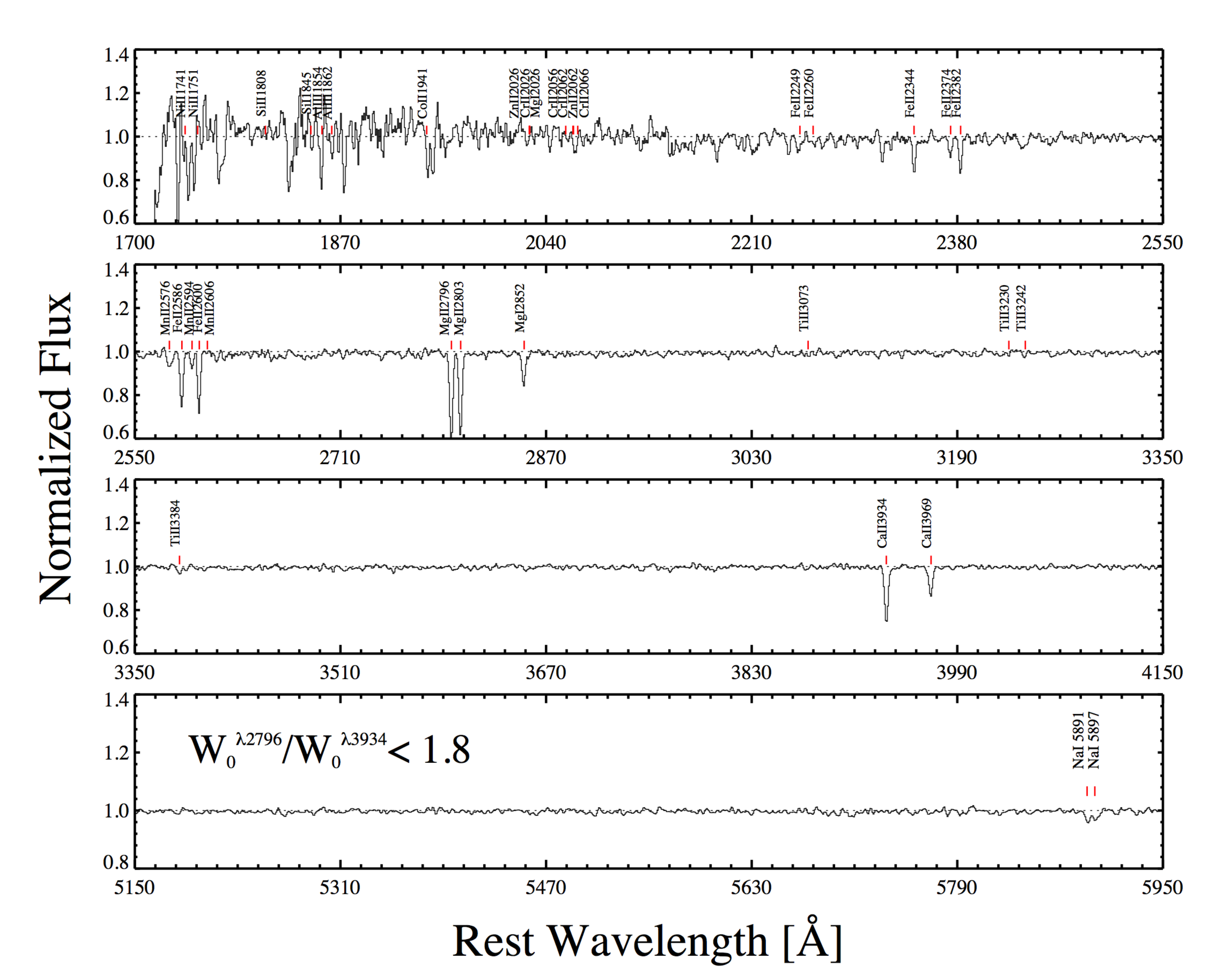} 
\caption{Same as Figure 2 but for the subsample of \CaII\ absorbers with W$_0^{\lambda2796}$/W$_o^{\lambda3934} < 1.8$.}
\label{Fig6} 
\end{figure*}

\begin{figure*} 
\includegraphics[width=0.95\textwidth]{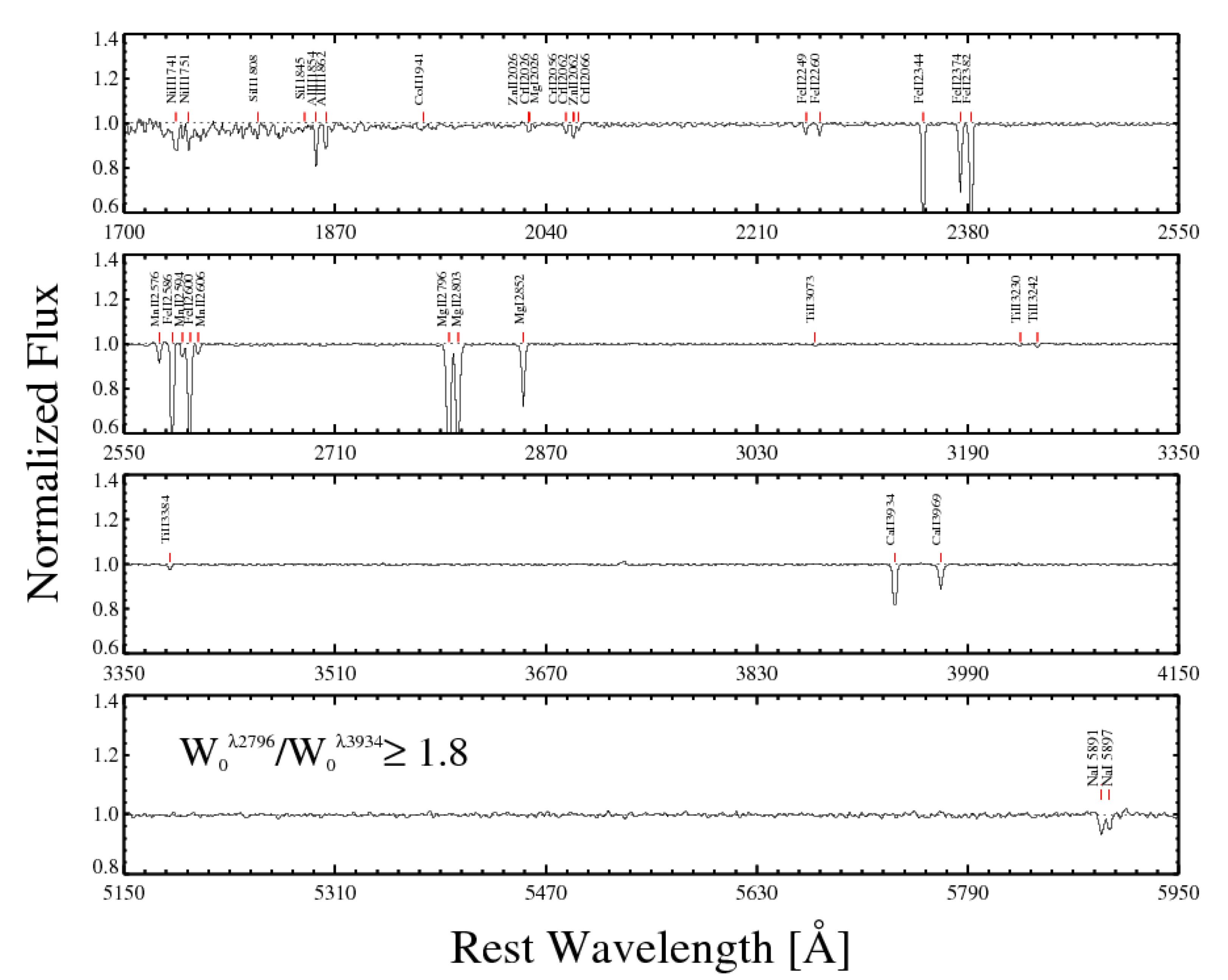} 
\caption{Same as Figure 2 but for the subsample of \CaII\ absorbers with W$_0^{\lambda2796}$/W$_o^{\lambda3934} \ge 1.8$.}
\label{Fig6} 
\end{figure*}

\input{TableOne}

\subsection{Column Densities and Element Abundance Ratios in \CaII\ Absorbers from their Composite Spectra}

Column densities derived using Eq. \ref{linearCOG} for the weak, unsaturated absorption lines in the full sample composite and the four subsample composites are reported in Table \ref{TableTwo} along with their 1$\sigma$ uncertainties. The reported results are generally variance-weighted averages of column densities determined from accessible transitions of the ion, similar to the results reported in \S3. 
We only provide results when significance levels are $\ge2 \sigma$, otherwise, 2$\sigma$ upper limits are reported.

\input{TableTwo}

The doublet ratios of \CaII\ and \NaI\ indicate that both may be partially saturated so we assign lower limits on their column densities. These column densities are consistent with those reported by previous authors using $\sim$10 times fewer systems (e.g., Wild et al. 2006, Nestor et al. 2008, Zych et al. 2009). However, none of these studies have sampled much of the $W_0^{\lambda 3934} \gtrsim 0.7$ \AA\ regime.  
% %Compare the values derived here versus other results.

As in \S3 we assume that the low-ionization column densities reported in Table \ref{TableTwo} represent the dominant ionization state due to self-shielding and that no significant ionization corrections are needed. The solar abundance ratios relative to Zn and Fe, i.e., [X/Zn] and [X/Fe], are then tabulated in Table \ref{TableThree} and Table \ref{TableFour}, respectively. 
%(xxx check to see if you want a separate table for abundance ratios relative to Fe. I assume you don't want a separate Table, you should say why it's useful in the next paragraph. Note that IF YOU DO WANT a table of abundances relative to Fe, you need to say why in the next paragraph. xxx
The ratios relative to Zn can reveal depletion of elements on to dust grains relative to solar abundances, while ratios relative to Fe can show important enhancements (see below). 

%with the exception of [Ti/Zn] which is rather consistent to the solar value xxx-I'm not sure what you are saying here, please clarify. Can we eliminate the 2nd sentence?-xxx. 

For the full sample, the results are generally consistent with DLA absorber populations over a range of redshifts (e.g., Turnshek et al. 1989, Pettini et al. 1999,  Prochaska \& Wolfe 2002, Ledoux et al. 2002, Prochaska et al. 2003, Akerman et al. 2005, Kulkarni et al. 2005, Battisti et al. 2012), 
%xxx note that I added a reference to myself since it was one of the first: Turnshek et al. 1989, ApJ, 344, 567 xxx, 
and with the ratios seen in individual \CaII~absorbers in the literature (e.g., Zych et al. 2009, Richter et al. 2011). 
Similar abundance ratios are also seen in metal-strong DLAs (MSDLAs), which are classified as those DLAs with logN(Zn$^+) \geq 13.15$ or $\log$N(Si$^+) \geq 15.95$ (Herbert-Fort et al. 2006), though the metal column densities of the \CaII\ absorbers are significantly lower than these values. The abundance ratios relative to Zn also approximately match the abundances ratios of the SMC as measured toward the star Sk 155 (Welty et al. 2001). 

The wavelength coverage of the \CaII\ absorber spectra permits the detection of both Fe-peak (e.g., Cr, Mn, Fe) 
and $\alpha$-capture elements (e.g., Si, Ca, Ti).\footnote{Ti is not an $\alpha$ element, but it shows abundance patterns similar to other $\alpha$ elements in Galactic stars (Edvardsson et al. 1995; Fran\c{c}ois et al. 2004)} It is generally thought that the Fe-peak and $\alpha$-capture elements are synthesized in the lead-up to Type II supernovae events over timescales $< 10^7$ years, whereas Fe-peak elements are also synthesized through Type Ia supernovae events occurring over $10^8 - 10^9 $ years. Hence, studying the abundance of $\alpha$-elements relative to Fe-peak elements provides clues to the chemical and star-formation patterns of the absorber. Furthermore, since different elements display various affinities to dust, one can also characterize the absorber depletion patterns. But disentangling the degeneracy between depletion and chemical enrichment is often difficult (e.g. Lauroesch et al. 1996, Lu et al. 1996, Prochaska \& Wolfe 2002, Vladilo 2002, Dessauges-Zavadsky, Prochaska \& D'Odorico 2002, Welty \& Crowther 2010). However, some constraints on the two effects can still be inferred from comparisons of various abundances against each other (Prochaska \& Wolfe 2002; Herbert-Fort et al. 2006). 
An enhanced [Ti/Fe] generally implies a Type II enrichment pattern, while an enhancement of [Si/Fe] 
suggests a population that is strongly depleted by dust. The [Ti/Zn] ratio is generally not a clear 
tracer of depletion, and using it would likely under-estimate the extent of depletion; however, 
the [Ti/Zn] ratio we observe for the \CaII\ absorbers does hint at some level of depletion of 
Ti on to dust grains. The enhancements of [Si/Fe], [Zn/Fe], and [Si/Ti] all unanimously 
indicate strong depletions in the typical \CaII\ absorber (Prochaska \& Wolfe 2002 ).

\input{TableThree}
\input{TableFour}

\subsection{Limits on Electron Densities in \CaII~Absorbing Gas }

The improvement in signal to noise which results when forming composite spectra allows us to derive some constraints on the electron density of the absorbing gas. Under ionization equilibrium, the balance between a neutral element $X$ and a singly ionized element $X^+$ is

\begin{equation} 
\frac{n(X^+)}{n(X^0)} = \frac{\Gamma}{\alpha(T) n_e}
\end{equation}

\noindent where n(X) denotes the volume density of X, $\Gamma$ is the photoionization rate of X to X$^{+}$, $\alpha$ is the temperature-dependent recombination coefficient to form X$^0$ from X$^+$,  and $n_e$ is the electron density. For gas of uniform density this can also be expressed using column densities by replacing $n(X)$ and $n(X^+)$ with the column densities $N(X)$ and $N(X^+)$, respectively (e.g., Prochaska, Chen \& Bloom 2006). In principle, constraints on $n_e$ could be obtained using observations of Fe, Ca, and  Mg transitions. However, except for \MgI, no transitions from a neutral atom are observed at $\geq 2\sigma$ in the composite. Also, the \CaII\ and \MgII\ doublet ratios show some indications of saturation. Therefore, the most conservative way to place constraints on $n_e$ is to use column density results derived from the observed \FeII\ lines and the absence of observed \FeI. These column density results are reported in Table \ref{TableTwo}. 

The frequency integral  of the product of the Fe photoionization cross section (Verner et al. 1996) and the local UV background (Mathis, Mezger, \& Panagia 1983) yields $\Gamma = 1.9 \times 10^{-10}$ s$^{-1}$. 
The total (radiative+dielectronic) recombination coefficient for 
Fe$^{+1}$ is $\alpha = 7.2 \times 10^{-13}$ cm$^3$ s$^{-1}$ at T$=10^4$ K (Mazzotta et al. 1998; Verner et al. 1999). 
This yields an average $2\sigma$ upper limit of $n_e \leq 2$ cm$^{-3}$ from the various 
composites, which is $\sim10$ times larger than the limits reported based on the analysis of 
higher-quality observations of individual absorbers by Nestor et al. (2008) and Zych et al. (2011).

\subsection{Dust in \CaII~Absorbing Gas} \label{red} 

The average extinction and reddening of \CaII\ absorbing gas can be derived by forming composites using the geometric mean of unnormalized flux spectra. This is done for the full sample of \CaII\ absorbers and the four subsamples.
The approach we take to form these composites is similar to the one taken in York et al. (2006). That is, when we construct a composite using \CaII\ absorber flux spectra, we also construct an unabsorbed reference composite.  

Specifically, we define a non-absorber match for every \CaII\ absorber in the sample. The match is determined using the quasar SDSS $i$-band magnitude and emission redshift $z_{em}$. To find a match, we formed a list of all SDSS quasars up to DR9 with $z_{em} > 0.1$ and $i < 20$ mag. We then eliminated those quasars with known intervening \MgII\ absorption (Quider et al. 2010; Monier et al., in prep, $private$ $communication$), \CaII\ absorption (Paper I), or  broad absorption lines (Shen et al. 2011). A tentative match between a \CaII\ absorber spectrum and an unabsorbed quasar spectrum is then determined by finding an unabsorbed quasar that lies closest to the absorbed quasar in $\Delta i$/$<\!\!i\!\!> - \Delta z_{em}$/$<\!\!z_{em}\!\!>$ space, where $\Delta i$ and $\Delta z_{em}$ represent differences between the absorbed and matching unabsorbed quasar properties.

\begin{figure*} 
\includegraphics[width=0.75\textwidth]{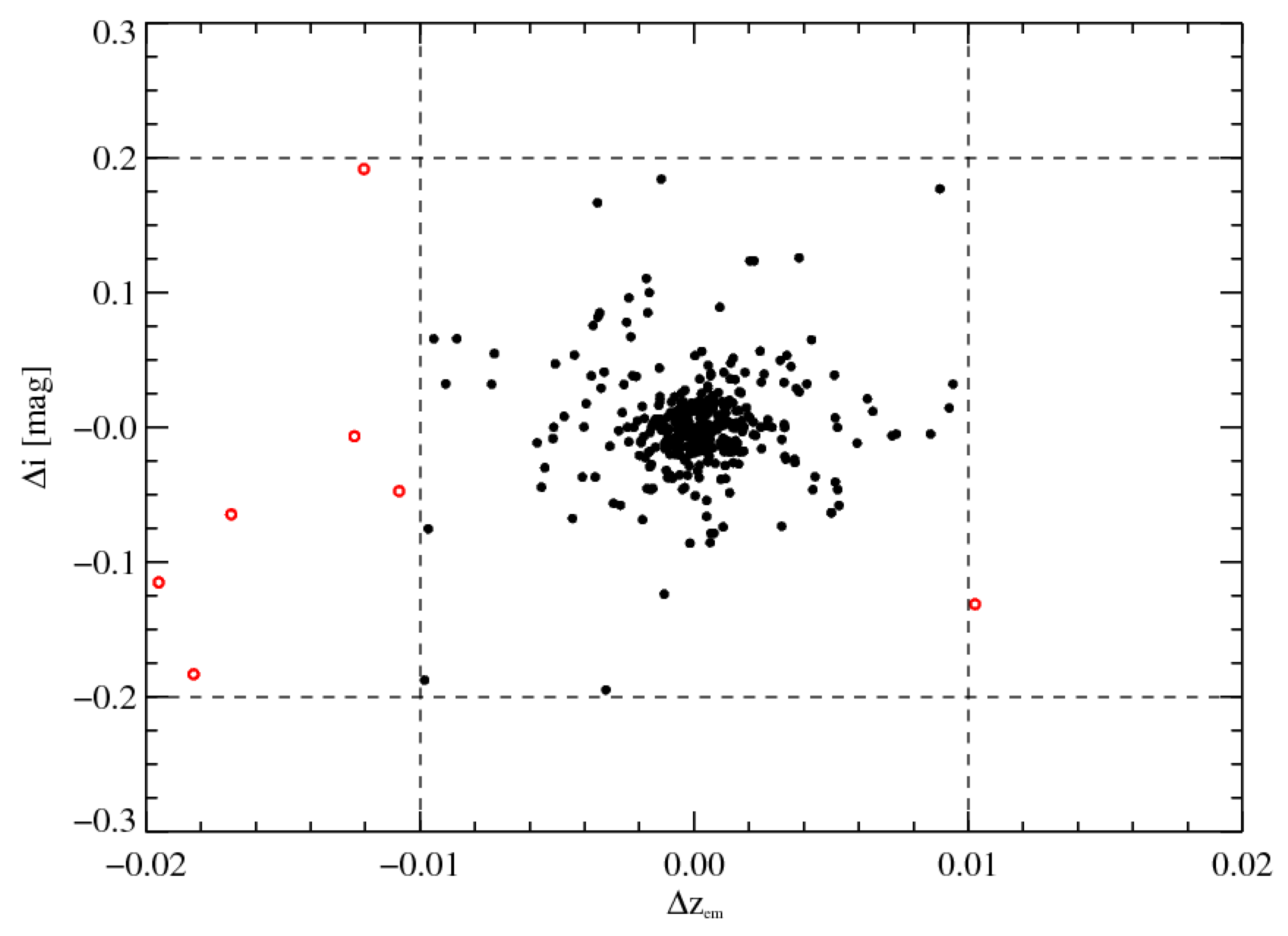} 
\caption{The $\Delta i$ - $\Delta z_{em}$ space for the final \CaII\ absorber and non-absorber matches. The dashed lines depict our initial match search box criteria. Suitable matches were found for 401 absorbers including the near matches shown as red open circles which we used.}
\label{MatchDist} 
\end{figure*}

\begin{figure*} 
\includegraphics[width=0.75\textwidth]{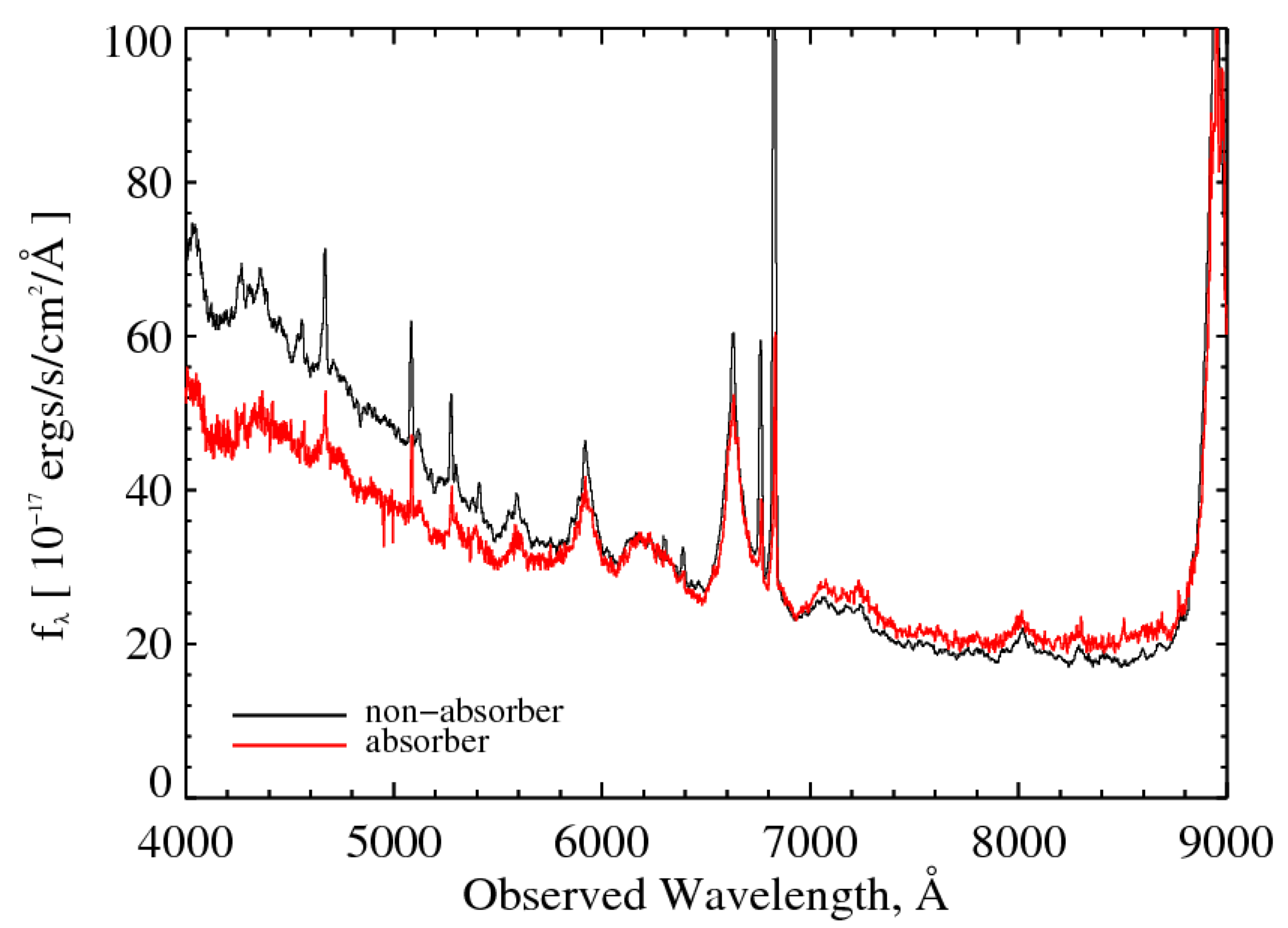} 
\caption{An example \CaII\ absorber spectrum and its non-absorber spectrum match. In this case $|\Delta i| = 0.01 $ and $|\Delta z_{em}| = 0.001$. }
\label{Match} 
\end{figure*}

Initial matches are accepted for those with $|\Delta i| \leq 0.20$ and $|\Delta z_{em}| \leq 0.01$ since we found it especially important to have similar emission features in both spectra. However, we also visually inspected initial matches for any missed absorption transitions because, of course, its presence in the unabsorbed list has the potential to cause extra extinction/reddening in the ``unabsorbed'' quasar spectrum, which could affect our analysis. In addition, we checked for other issues such as broad intrinsic \FeII\ emission in the matched unabsorbed quasar spectra, which may lead to significant unmatched broad features in individual absorbed and unabsorbed spectra. 
%Evidently, such issues could result to features in the composite spectrum that might also confuse the interpretation of results. 
When these types of problems occurred, we removed the quasar from the unabsorbed quasar list and re-ran the matching process until a satisfactory match was found. In the end, we found that it was not possible to find a suitable match with $|\Delta i| \leq 0.2$ and $|\Delta z_{em}| \leq 0.01$ for $41$ of our \CaII\ absorbers, which is $\sim10$\% of our full sample. We did accept seven of those $41$ since the matches were not too discrepant. Figure \ref{MatchDist} illustrates the distribution of $\Delta i$ and  $\Delta z_{em}$ for our final matches. Red points mark the seven closer outliers which just missed matching our search criteria but were included. Figure \ref{Match} illustrates an example case of an individual \CaII\ absorber and its match. Fifty per cent of the final matches are within $|\Delta i| \leq 0.014$ mag and $|\Delta z_{em}| \leq 0.001$, and the median and average values of the distributions of $\Delta i$ and $\Delta z_{em}$ are indistinguishable from zero.

%As in York et al. (2006), 
To assess extinction and reddening we are interested in comparing the \CaII\ absorber fluxed composite continuum (which is in the rest frame of the absorber) to the unabsorbed reference continuum. To this end, 
we constructed the composites for our extinction analysis by taking 
the geometric mean of those \CaII\ absorbed spectra which have suitable unabsorbed matches, hence we used 401 sets of absorbed-unabsorbed spectra. A quasar continuum generally follows a power-law and the geometric mean of a set of power-law spectra preserves the average power-law index. Therefore, this is an appropriate method to use to determine the average extinction law, which is also likely to be similar to a power-law. 

\begin{figure*} 
\includegraphics[width=0.95\textwidth]{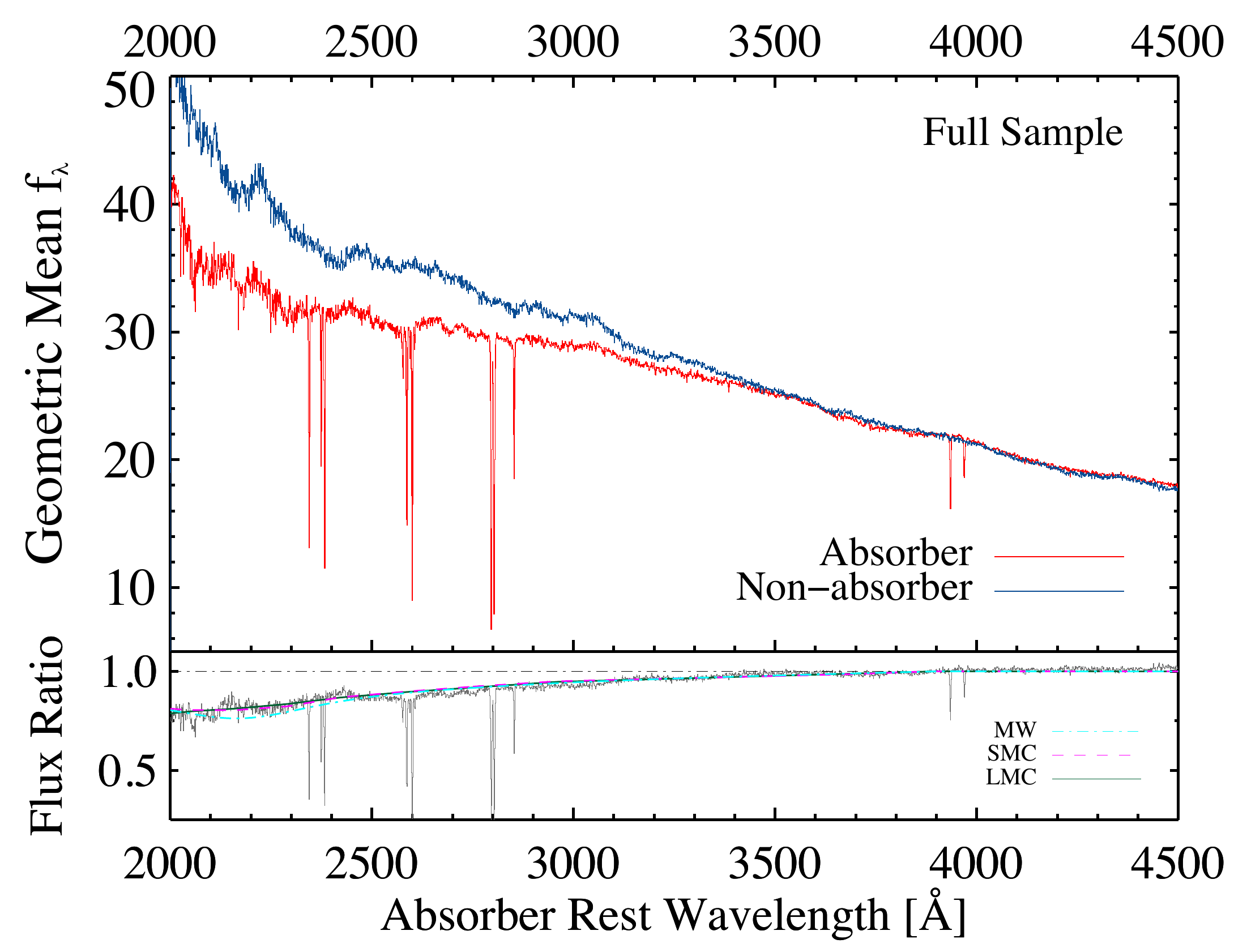} 
\caption{\textit{Top:} The geometric mean rest-frame composites for the \CaII\ absorbers (red) and the unabsorbed reference sample (blue) derived using the full sample. Clearly visible in the absorber composite are the narrow absorption lines from \FeII, \MgII, \MgI, and \CaII. \textit{Bottom:} The ratio of the absorber composite to the unabsorbed reference composite. Least-squares fits of dust models derived from the LMC (dashed line in magenta), SMC (solid line in green), and Milky Way (MW) (dashed-dot line in cyan) are also shown. The LMC and SMC dust laws both provide good fits to the observe extinction, with $E(B-V) \approx 0.03$. The LMC and SMC fits are nearly indistinguishable.} %Why The emission lines are spread out %: They are smeared out because of the spread in the dritibution of z_em - zabs}
\label{Extinct400} 
\end{figure*}

To form a fluxed composite each \CaII\ absorbed quasar spectrum and its match were shifted to the absorber rest frame after rebinning to a wavelength scale that was one-tenth of the original pixel size in order to make the registration of spectra accurate.
%, even though there were no absorbers in the reference spectra. 
%We used finer grid sizes that were about a tenth of the original size. Here, we also assume that spectroscopic properties of thequasar in the absorber and non-absorber samples are statistically equivalent, except for the effects of the absorption systems. 
The final composites were then rebinned to a wavelength scale of 1 \AA\ per pixel in the rest frame. 
The standard deviation, $\sigma_{\lambda,g}$, in the geometric mean is given by

\begin{equation} 
\ln\sigma_{\lambda,g} =  \sqrt{\frac{1}{N}\sum_{i=1}^{N} \left[\ln{\left(\frac{f_{\lambda, i}}{\mu_{\lambda,g}}\right)}\right]^2 } 
\end{equation}

\noindent where $\mu_{\lambda,g}$ is the geometric mean of $N$ fluxes and $f_{\lambda, i}$ are the individual spectrum fluxes as a function of wavelength (Kirkwood 1979). The top panel of Figure \ref{Extinct400} shows the matched composites for the full sample, with the \CaII\ absorber sample in red and the unabsorbed reference sample in blue. Figures 10-11 illustrate this in the four subsamples (see \S5 discussion.) The errors in the full sample composite typically lie in the range  $1.8 - 2.2 \times 10^{-17}$ ergs cm$^{-2}$ s$^{-1}$ \AA$^{-1}$ level. Prominent absorption features from transitions of \FeII, \MgII, \MgI, and \CaII\ are clear in the \CaII\ absorber composite. The bottom panel shows the flux ratios between the absorbed and unabsorbed composites. Overlaid are extinction model fits to the data: the magenta dashed line shows a LMC-like dust model (Gordon et al. 2003), the solid green line shows a SMC-like model (Gordon et al. 2003), and the dashed-dot cyan lines show a standard Milky Way model (Fitzpatrick 1999). All fits have been made over a wavelength range $\geq 2500$ \AA\ to ensure more uniform noise characteristics across the wavelength range when performing the fits. 

In Table \ref{TableZero} we summarize the modeled or observed absorbed-to-unabsorbed flux ratios at $\lambda_{rest} = 2200$ \AA, $\mathcal{R}$, and color excesses, $E(B-V)$, for the full sample and four subsamples. For the full sample the observed $\mathcal{R}$ is 0.83, which is best matched by either the LMC or SMC; the color excess is inferred to be $E(B-V) \approx 0.03$.  Over the fitted range ($\lambda > 2500$ \AA) the LMC and SMC models are nearly indistinguishable, while the MW dust law is definitely ruled out.
%Uncertainties in the color excess values are at $\sim10$\%. 
Wild et al. (2006) concluded that an LMC extinction law applied to the \CaII\ sample of absorbers that they studied. We also note for comparison that an SMC-like extinction curve with $E(B-V) < 0.02$ mag has been inferred using DLA and \MgII\ samples (Murphy \& Liske 2004; York et al. 2006; Vladilo, Prochaska \& Wolfe 2008). 

\begin{figure*} 
\includegraphics[width=0.95\textwidth]{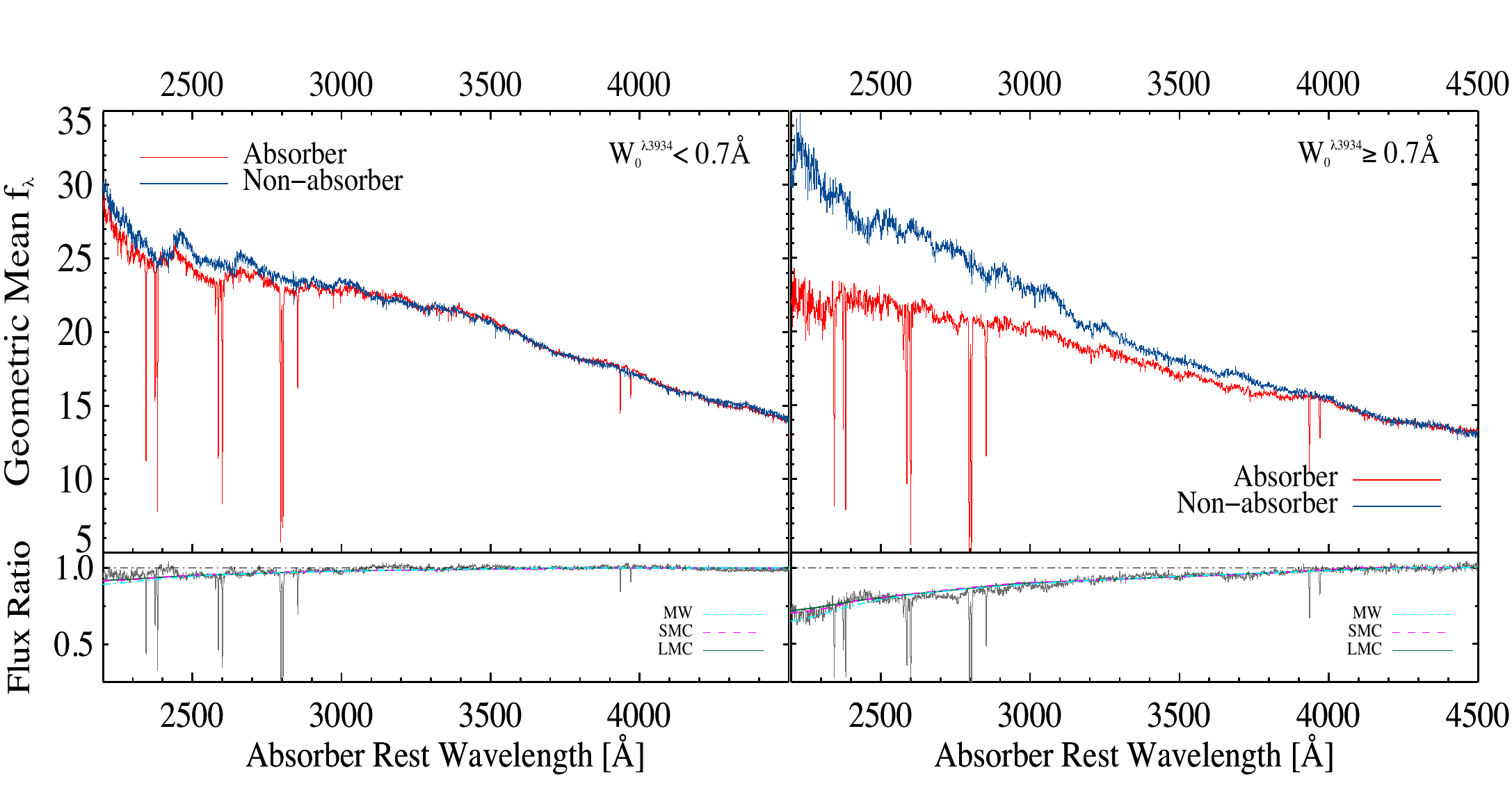} 
\caption{Same as Figure 9, but showing the results for the W$_o^{\lambda3934} < 0.7$ \AA\ subsample (left side) and W$_o^{\lambda3934} \ge 0.7$ \AA\ subsample (right side).}
\label{Fig10} 
\end{figure*}

\begin{figure*} 
\includegraphics[width=0.95\textwidth]{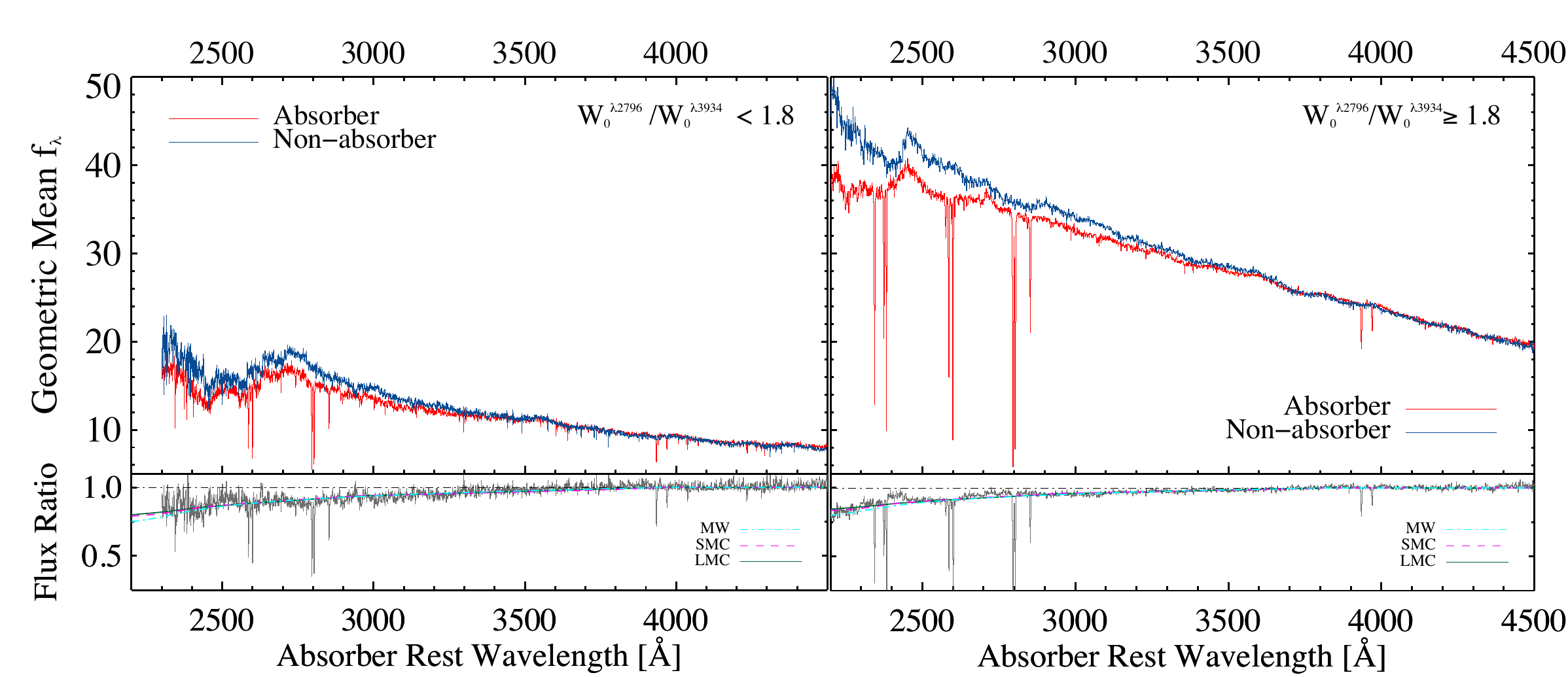} 
\caption{Same as Figure 9, but showing the results for the W$_o^{\lambda2796}$/W$_o^{\lambda3934} < 1.8$ subsample (left side) and W$_o^{\lambda2796}$/W$_o^{\lambda3934} \ge 1.8$ subsample (right side).}
\label{Fig11} 
\end{figure*}

\section{Implications of the Results for \CaII\ Absorber Populations Using the Subsamples}

As indicated at the beginning of \S4, in Paper I we found that while \CaII\ absorbers are rare, they are unlikely to represent a single type or population of absorber. The W$_0^{\lambda3934}$ distribution requires a two-component exponential to satisfactorily fit the data, hinting at the existence of at least two distinct populations.
% of \CaII\ absorbers. It has also been demonstrated that such a two-component distribution appears to 
This persists across our survey redshift interval, $z_{abs} \lesssim 1.4$.  Upon further analysis of the \CaII\ survey data, it was also shown that when the \MgII\ properties of these \CaII\ absorbers are taken into account, it is possible to more clearly separate the \CaII\ absorbers into two populations at the $> 99\%$ confidence level. In this section we investigate whether the chemical and dust depletion properties of subsamples of \CaII\ absorbers are consistent with the statistical evidence for two populations.

To do this, we divide the full sample into four subsamples, and we analyze the subsamples in the same way we analyze the full sample as discussed in \S4. The tabulations of results and the figures on subsamples are in Tables  $3-7$ and Figures $3-6$ and $10-11$. Recall (Paper I and \S4) that we divide the full sample as follows. Two subsamples were formed by separating the full sample at $W_0^{\lambda3934} = 0.7$ \AA, which results in $\sim 200$ \CaII\ absorbers in each. This is also the separation which exhibits the maximum difference between two populations from KS tests. We also form two more subsamples by separating the full sample at $W_0^{\lambda2796}$/$W_0^{\lambda3934} = 1.8$, but only $\sim 30$ \CaII\ absorbers have $W_0^{\lambda2796}$/$W_0^{\lambda3934} < 1.8$. 

Figures \ref{Fig3} and \ref{Fig4} show the normalized composite spectra of the two subsamples of \CaII\ absorbers separated at $W_0^{\lambda3934} = 0.7$ \AA. %As in the \S3, features with $\geq 2\sigma$ significance are labeled in red. The emission line [\OII] is marked in blue. 
The measurements of the equivalent widths are reported in Table \ref{TableOne}. The corresponding ionic column densities derived from unsaturated lines of \CrII, \ZnII, \FeII, \NiII, and \MnII\ are tabulated in Table \ref{TableTwo}. Estimates on the abundance ratios relative to Zn and Fe are inferred in Tables \ref{TableThree} and \ref{TableFour}, assuming no ionization corrections. The results clearly indicate that the two subsamples separated at $W_0^{\lambda3934} = 0.7$ \AA\ reveal the existence of two broadly defined populations of \CaII\ absorbers in terms of their element abundance ratios and depletion measures, although there is likely some cross-mixing between the two populations given the crude way they were separated. However, the two subsamples formed by separating the full sample at $W_0^{\lambda2796}$/$W_0^{\lambda3934} = 1.8$ do not allow us to draw a similar type of conclusion because only $\sim 30$ \CaII\ absorbers have $W_0^{\lambda2796}$/$W_0^{\lambda3934} < 1.8$, which compromises the accuracy of this particular measurement. Results from the $W_0^{\lambda2796}$/$W_0^{\lambda3934} \ge 1.8$ subsample are very similar to $W_0^{\lambda3934} < 0.7$ \AA\ results. 
 
\begin{figure*} 
\includegraphics[width=0.75\textwidth]{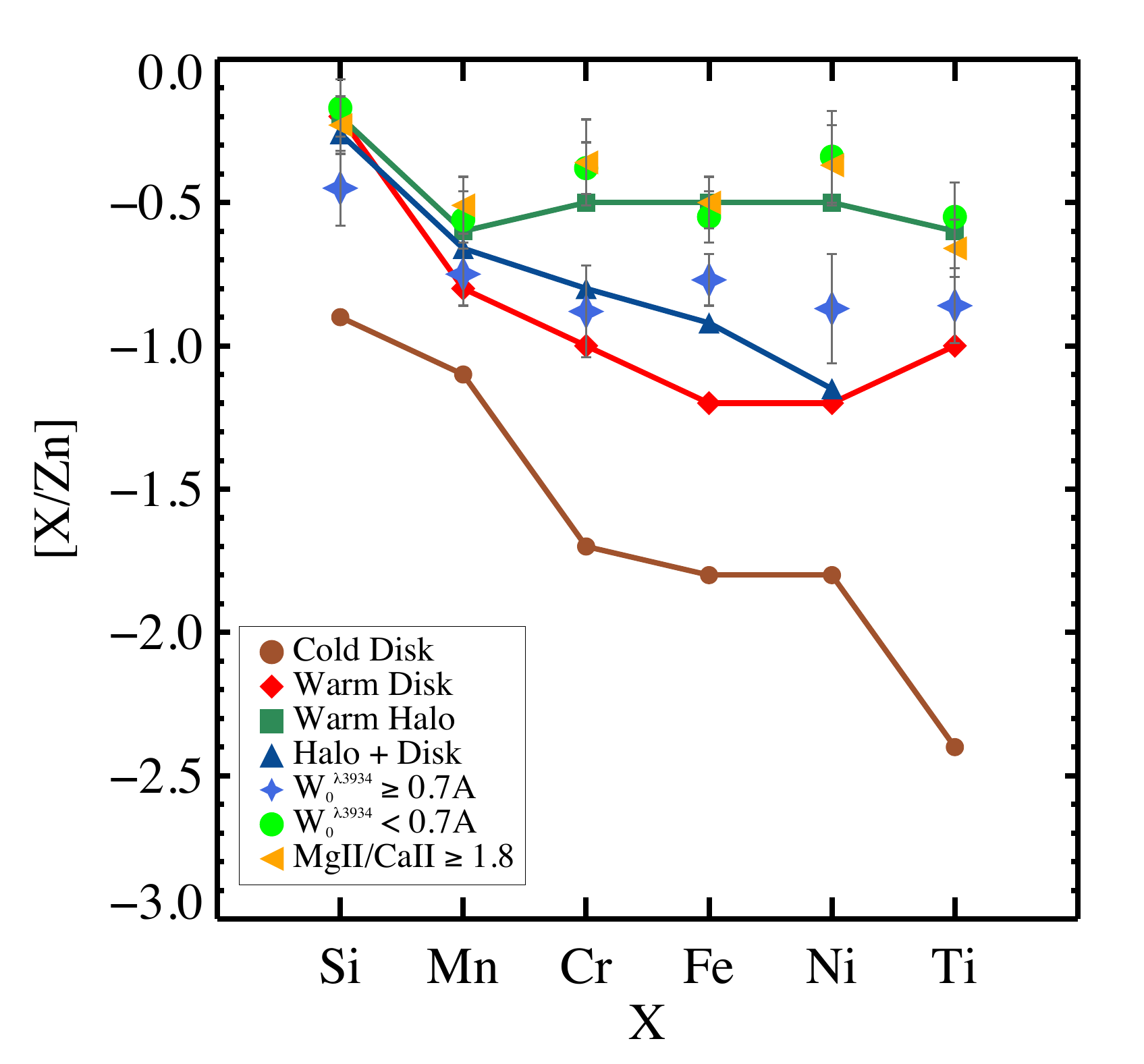} 
\caption{Element abundance ratios relative to Zn for three \CaII\ absorbers subsamples: (1) W$_0^{\lambda3934} < 0.7$ \AA, (2) W$_0^{\lambda3934} \ge 0.7$ \AA, (3) and W$_0^{\lambda2796}$/W$_0^{\lambda3934} \ge 1.8$. The \CaII\ absorber subsample with W$_0^{\lambda2796}$/W$_0^{\lambda3934} < 1.8$ is not shown because of the small number of absorbers (and large error bars) that pertain to this subsample. The elements are arranged in order increasing condensation temperature. For comparison, and as described in the text, also shown are abundance ratio compilations for cold disk gas, warm disk gas, disk + halo gas, and warm halo gas (Welty et al. 1999; Savage \& Sembach 1996).}
\label{CoolWarmDisk} 
\end{figure*}

Figure \ref{CoolWarmDisk} shows the log of the abundance ratios of Si, Mn, Cr, Fe, Ni, and Ti relative to Zn, as measured with respect to solar values. The elements on the x-axis are ordered left-to-right in increasing condensation temperatures. Filled symbols are Galactic abundance measurements for the halo (green squares), cold disk (brown circles), and warm disk (red diamonds) obtained from the compilations of Welty et al. (1999). The halo + disk (blue triangles) abundances are taken from  Savage \& Sembach (1996). Measurements pertaining to \CaII\ absorbers with 
$W_0^{\lambda3934} < 0.7$ \AA\ are shown as filled green circles while those with $W_0^{\lambda3934} \geq 0.7$ \AA\ are blue asterisks; results for \CaII\ absorbers with $W_0^{\lambda2796}$/$W_0^{\lambda3934} \ge 1.8$ are shown as filled orange triangles and are seen to be very similar to the $W_0^{\lambda3934} < 0.7$ \AA\ results; the $W_0^{\lambda2796}$/$W_0^{\lambda3934} < 1.8$ results are not shown due to their poor accuracy.

Figures 10 and 11 illustrate the extinction and reddening results for the four subsamples.  A tabulation of observed results and best-fit extinction laws are given in Table 7.

The stronger \CaII\ absorber subsample ($W_0^{\lambda3934} \ge 0.7$ \AA) is seen in Figure \ref{CoolWarmDisk} to be similar to the halo + disk component in terms of both chemical enrichment and element depletions on to dust grains. This conclusion is consistent with both the best-fit LMC or SMC dust extinction laws for our stronger \CaII\ absorber subsample (right panel of Figure 10). This is our most heavily reddened subsample, with an absorbed-to-unabsorbed flux ratio at 2200 \AA\ of $\mathcal{R}=0.73$ (Table 7). A MW extinction law is clearly ruled out for the stronger \CaII\ absorber subsample.  
However, the weaker \CaII\ absorber subsample ($W_0^{\lambda3934} < 0.7$ \AA) is seen to have chemical enrichment and element depletion characteristics of the warm halo component, and there is much less reddening due to dust extinction (left panel of Figure 10 and Table 7), with $\mathcal{R}=0.95$. Calculating the optical depths from the flux ratios indicate that the stronger \CaII\ absorbers are nearly six times more reddened than the weaker absorbers.

Previous studies of the stronger \CaII\ absorption systems have been very limited until now. In the future it will be important to explore the properties of individual \CaII\ absorption systems (especially the stronger ones) with high-resolution spectroscopy in order to measure their kinematics and better characterize their chemistry.

\input{TableZero}

\section{Summary and Conclusions}
%\textcolor{red}{I will have to include preliminary results from reddening like we discussed.}

We have used statistical results on the 435 \CaII\ absorbers identified in SDSS quasar spectra (Paper I) to derive results on their element abundance ratios and dust properties.
%Over 400 \CaII\ absorbers were compiled by Sardane et al. (2014) in Paper I. 
%from DR7+DR9 quasars with $i < 20$, using selection cuts of 5$\sigma$-2.5$\sigma$ W$_0^{\lambda3934}$ and W$_0^{\lambda3969}$ and physical doublet ratios. 
In contrast to earlier studies, this new large sample includes a large number ($\approx 200$) of $W_0^{\lambda3934} \geq 0.7$ \AA\ \CaII\ absorbers at redshifts $z_{abs} < 1.4$. We present results on a number of individual \CaII\ absorption systems in Tables 1 and 2. More importantly, by median-combining $>$ 400 normalized spectra for the full sample and four subsamples, we have formed high signal-to-noise normalized composite spectra and used them to detect (or place limits on) low-ionization metal lines due to \SiII, \FeII, \TiII, \CoII, \ZnII, \CrII, \FeI, \SiI, \MnII, \MgII, \MgI, \NiII, \TiII, \CaII\ and \NaI, as included within the redshifted spectral coverage of the SDSS spectrograph. These have been used to investigate element abundance ratios in \CaII\ absorbers. We also formed \CaII\ absorber fluxed composite spectra and matching unabsorbed fluxed composite spectra of the full sample and four subsamples to investigate extinction and reddening in \CaII\ absorbers. 

%A weak [\OII] emission is also detected, a star-formation indicator, but due to the low signal-to-noise ratio of the detection, we are not able to place a much tighter constraint on the upper limit of the electron number density than what has been available in the literature (e.g. Zych et al. 2009). The metal-line column densities in the median stack are consistent with results found by previous authors having $\lesssim$10 times smaller sample sizes. 

We tested a hypothesis put forth in Paper I. Namely,  that the sensitivity-corrected W$_0^{\lambda3934}$ distribution of \CaII\ absorbers follows a shape that is suggestive of at least two populations of \CaII\ absorbers, separated at  W$_0^{\lambda3934} = 0.7$ \AA. We therefore hypothesized that analysis of two subsamples divided at W$_0^{\lambda3934} = 0.7$ \AA\ would allow us to reveal the nature of these two populations, and this turned out to be the case. We also showed in Paper I that by using information on \MgII\ in \CaII\ absorbers we could statistically infer the presence of two populations divided at  W$_0^{\lambda2796}$/W$_0^{\lambda3934} = 1.8$, but unfortunately only $\sim30$ \CaII\ absorbers are in the subsample with W$_0^{\lambda2796}$/W$_0^{\lambda3934} < 1.8$, so we could not make an accurate comparison of these other two subsamples using this criterion.  

Because of our findings, in what follows we will refer to the W$_0^{\lambda3934} \ge 0.7$ \AA\
absorbers as the strong \CaII\ absorbers, and the W$_0^{\lambda3934} < 0.7$ \AA\ absorbers as the weak \CaII\ absorbers.

Analysis of the element abundance ratios derived for Si, Mn, Cr, Fe, Ni, and Ti relative to Zn using normalized composite spectra indicate that the abundance pattern of the strong \CaII\ absorbers is intermediate between  disk- and halo-type gas (see Figure 12). The results indicate more significant depletions of the highly refractory elements of Cr, Fe, Ni, and Ti in the strong \CaII\ absorbers. In addition, independent of the absorption line analysis, which was based on normalized composites, an investigation of the extinction and reddening in the strong \CaII\ absorbers using the ratio of the absorbed-to-unabsorbed composite fluxed spectra shows that they are a significantly reddened population of absorbers, with the absorbed-to-unabsorbed composite flux ratio at $\lambda_{rest} = 2200$ \AA\ being $\mathcal{R} \approx 0.73$ and $E(B-V) \approx 0.046$, consistent with a LMC or SMC dust law (right hand panel of Figure 10 and Table 7). Our data do not allow us to distinguish between an LMC versus SMC reddening law. 

At the same time, we showed that the weak \CaII\ absorbers have an abundance pattern typical of halo-type gas with less depletion of the highly refractory elements of Cr, Fe, Ni, and Ti (also Figure 12). Again independent of the absorption line analysis, we find that the weak \CaII\ absorbers are nearly six times less reddened than the strong \CaII\ absorbers, with $\mathcal{R} \approx 0.95$ and $E(B-V) \approx 0.011$ (left hand panel of Figure 10 and Table 7).

Thus, the results of this analysis have confirmed the hypothesis that at least two populations of \CaII\ absorbers exist, consistent with the statistical evidence in Paper I. 
%picture of the multi-population scenario statistically inferred from the $W_0^{\lambda3934}$ distribution. Splitting the population of \CaII\ absorbers into two roughly equal components at , indicate 
Thanks to the high-signal-to-noise composite spectra, we were able to identify the striking differences in the element abundance ratios, depletion patterns,  and dust extinction and reddening properties of the two populations of \CaII\ absorbers divided at W$_0^{\lambda3934} = 0.7$  \AA.
In Paper III we will explore the association between \CaII\ absorbers and galaxies.

%\newpage

\nocite{*} 
\bibliography{ReferencesPaper2}{} 
\bibliographystyle{mn2e}
\section*{Acknowledgments}

GMS acknowledges support from a Zaccheus Daniel Fellowship and a Dietrich School of Arts and
Sciences Graduate PITT PACC Fellowship from the University of Pittsburgh.

Funding for SDSS has been provided by the 
Alfred P. Sloan Foundation, the Participating Institutions, the National Science Foundation, 
and the US Department of Energy Office of Science. The SDSS is managed by the Astrophysical 
Research Consortium for the Participating Institutions. The Participating Institutions are 
the American Museum of Natural History, Astrophysical Institute Potsdam, 
University of Basel, University of Cambridge, Case Western Reserve University, University of Chicago, 
Drexel University, Fermilab, the Institute for Advanced Study, the Japan Participation Group, Johns Hopkins 
University, the Joint Institute for Nuclear Astrophysics, the Kavli Institute for Particle Astrophysics 
and Cosmology, the Korean Scientist Group, the Chinese Academy of Sciences (LAMOST), Los 
Alamos National Laboratory, the Max-Planck-Institute for Astronomy (MPIA), the 
Max-Planck-Institute for Astrophysics (MPA), New Mexico State University, Ohio State University, 
University of Pittsburgh, University of Portsmouth, Princeton University, 
the United States Naval Observatory, and the University of Washington. 

\end{document}

%% file: TableFive.tex
\begin{table*}
%\small
\def\arraystretch{1.5}

%\caption{\ZnII, \CrII, \FeII~and \MnII~abundances of \CaII~absorbers with $z_{abs} \gtrsim 0.8$, derived from the equivalent widths of detectable weak transitions. The column densities are taken to be the error-weighted means of the column densities derived from various transitions of the same ionic species. }
\caption{Cr$^{+1}$, Zn$^{+1}$, Fe$^{+1}$, and Mn$^{+1}$ column densities of \CaII~absorbers derived from the equivalent widths of detectable weak transitions. The \CaII\ absorbers all lie in the redshift interval $0.87 \lesssim z_{abs} \lesssim 1.21$. The column densities of the various low-ionization elements are derived as described in the text.}

\begin{tabular}{lc cc cccc}
\hline
Quasar & $z_{abs}$ & $W_0^{\lambda3934}$ & $W_0^{\lambda3969}$ &  log $N$(Cr$^{+1}$) & log $N$(Zn$^{+1}$) & log $N$(Fe$^{+1}$)  & log $N$(Mn$^{+1}$)]   \\
                &           & ($\mathrm{\AA}$)      & ($\mathrm{\AA}$)      &  \multicolumn{4}{c|} { (atoms  cm$^{-2}$)} \\

\hline 
J081053+352224 & 0.877 & 0.509 $\pm$ 0.074 & 0.254 $\pm$ 0.078 & 13.30 $\pm$ 0.15 &  12.84 $\pm$  0.10 & 14.61 $\pm$ 0.04 & 12.31 $\pm$  0.04 \\
J114658+395834 & 0.900 & 0.381 $\pm$ 0.042 & 0.137 $\pm$ 0.045 & 13.86 $\pm$ 0.02 &  12.09 $\pm$  0.02 & 14.23 $\pm$ 0.02 & 11.93 $\pm$  0.02 \\
J153503+311832 & 0.904 & 0.524 $\pm$ 0.045 & 0.387 $\pm$ 0.044 & 12.87 $\pm$ 0.08 &  11.90 $\pm$  0.12 & 14.16 $\pm$ 0.07 & 11.84 $\pm$  0.04 \\
J162558+313911 & 0.906 & 0.813 $\pm$ 0.156 & 0.332 $\pm$ 0.130 & 13.63 $\pm$ 0.07 &  12.58 $\pm$  0.10 & 14.96 $\pm$ 0.27 & 12.46 $\pm$  0.04 \\
J100000+514416 & 0.907 & 0.896 $\pm$ 0.168 & 0.660 $\pm$ 0.216 & 13.34 $\pm$ 0.21 &  12.59 $\pm$  0.12 & 14.53 $\pm$ 0.40 & 12.36 $\pm$  0.13 \\
J094145+303503 & 0.938 & 1.118 $\pm$ 0.095 & 0.872 $\pm$ 0.104 & 13.45 $\pm$ 0.12 &  12.34 $\pm$  0.06 & 14.62 $\pm$ 0.08 & 12.21 $\pm$  0.03 \\
J172739+530229 & 0.945 & 0.590 $\pm$ 0.094 & 0.422 $\pm$ 0.112 & 13.43 $\pm$ 0.13 &  12.49 $\pm$  0.19 & 14.50 $\pm$ 0.06 & 12.15 $\pm$  0.04 \\
J112932+020422 & 0.965 & 0.632 $\pm$ 0.051 & 0.489 $\pm$ 0.063 & 13.09 $\pm$ 0.09 &  11.97 $\pm$  0.41 & 14.30 $\pm$ 0.04 & 11.93 $\pm$  0.04 \\
J233917-002943 & 0.967 & 0.475 $\pm$ 0.095 & 0.439 $\pm$ 0.111 & 13.62 $\pm$ 0.08 &  12.42 $\pm$  0.18 &  \nodata         & 12.39 $\pm$  0.07 \\
J014717+125808 & 1.039 & 0.484 $\pm$ 0.065 & 0.253 $\pm$ 0.066 & 13.36 $\pm$ 0.08 &  12.08 $\pm$  0.05 & 14.40 $\pm$ 0.04 & 12.07 $\pm$  0.03 \\
J213408+043611 & 1.118 & 0.804 $\pm$ 0.085 & 0.426 $\pm$ 0.089 & 12.68 $\pm$ 0.13 &  11.96 $\pm$  0.06 & 14.28 $\pm$ 0.08 & 11.97 $\pm$  0.11 \\
J141615+365537 & 1.204 & 0.696 $\pm$ 0.086 & 0.396 $\pm$ 0.063 & 13.19 $\pm$ 0.11 &  11.89 $\pm$  0.27 &  \nodata         & 12.09 $\pm$  0.83 \\
\hline

\end{tabular}
\label{TableFive}
\end{table*}

%0429-51820-0215
%0892-52378-0106
%0385-51877-0229
%4730-55630-0120
%0359-51821-0042
%1942-53415-0545
%0903-52400-0258
%1684-53239-0040
%4722-55735-0584
%4654-55659-0886
%3855-55268-0942
%4084-55447-0090

%% file: TableSix.tex
\begin{table*}
%\small
\def\arraystretch{1.5}
\caption{Measurements of \NaI\ $\lambda\lambda$5891,5897 doublet rest equivalent widths for \CaII~absorbers 
with $z_{abs} \lesssim 0.7$. The table is ordered in terms of increasing $z_{abs}$. For weak unsaturated lines, we derive the Na$^+$ column densities as described in the text. For \NaI\ profiles with doublet ratios approaching 
saturation, lower limits on N(Na$^+$) are reported. }

\begin{tabular}{c ccc c cccc c}
\hline
Quasar    & $z_{abs}$ &  $W_0^{\lambda 3934}$ &  $W_0^{\lambda 3969}$ &  $W_0^{\lambda 5891}$ & $W_0^{\lambda 5897}$ & log $N$(Na$^{+1}$)\\
          &           &  ($\mathrm{\AA}$)       &   ($\mathrm{\AA}$)      &   ($\mathrm{\AA}$)      &   ($\mathrm{\AA}$)     & (atoms cm$^{-2}$)\\
\hline
J155752+342140 &	0.114	&	0.598	$\pm$	0.102	&	0.628	$\pm$	0.168	&	0.527	$\pm$	0.165	&	0.380	$\pm$	0.154	&	$\geq$	12.54\\	
J075031+192754 &	0.180	&	0.437	$\pm$	0.084	&	0.447	$\pm$	0.098	&	0.580	$\pm$	0.123	&	0.380	$\pm$	0.097	&	$\geq$	12.55\\	
J091958+111152 &	0.182	&	1.105	$\pm$	0.212	&	0.720	$\pm$	0.168	&	0.407	$\pm$	0.138	&	0.569	$\pm$	0.144	&	$\geq$	12.66\\	
J085917+105509 &	0.183	&	0.92	$\pm$	0.108	&	0.431	$\pm$	0.115	&	0.277	$\pm$	0.098	&	0.209	$\pm$	0.103	&	$\geq$	12.28\\	
J114339+073105 &	0.189	&	0.632	$\pm$	0.098	&	0.462	$\pm$	0.080	&	0.420	$\pm$	0.089	&	0.180	$\pm$	0.095	&	12.29	$\pm$	0.09\\
J142536-001702 &	0.220	&	1.111	$\pm$	0.094	&	0.515	$\pm$	0.079	&	0.181	$\pm$	0.091	&	0.150	$\pm$	0.093	&	$\geq$	12.13\\	
J085045+563618 &	0.225	&	0.532	$\pm$	0.071	&	0.219	$\pm$	0.073	&	1.226	$\pm$	0.121	&	1.068	$\pm$	0.099	&	$\geq$	12.96\\	
J082312+264415 &	0.253	&	0.633	$\pm$	0.103	&	0.383	$\pm$	0.102	&	0.673	$\pm$	0.165	&	0.398	$\pm$	0.154	&	12.59	$\pm$	0.08\\
J165743+221149 &	0.266	&	1.642	$\pm$	0.221	&	1.546	$\pm$	0.158	&	1.838	$\pm$	0.176	&	1.060	$\pm$	0.200	&	$\geq$	13.02\\	
J124300+204246 &	0.277	&	1.488	$\pm$	0.126	&	1.034	$\pm$	0.101	&	1.716	$\pm$	0.145	&	0.957	$\pm$	0.120	&	$\geq$	12.97\\	
J085010+593118 &	0.282	&	0.279	$\pm$	0.043	&	0.160	$\pm$	0.044	&	0.310	$\pm$	0.054	&	0.194	$\pm$	0.067	&	$\geq$	12.27\\	
J102935-012138 &	0.290	&	0.351	$\pm$	0.067	&	0.208	$\pm$	0.055	&	0.177	$\pm$	0.084	&	0.240	$\pm$	0.118	&	$\geq$	12.31\\	
%J094927+314110 &	0.305	&	1.779	$\pm$	0.056	&	1.065	$\pm$	0.062	&	25.420	$\pm$	0.256	&	22.964	$\pm$	0.279	&	$\geq$	14.10\\	
J152800+535223 &	0.316	&	0.541	$\pm$	0.056	&	0.325	$\pm$	0.080	&	0.475	$\pm$	0.147	&	0.186	$\pm$	0.108	&	12.33	$\pm$	0.13\\
J161649+415416 &	0.321	&	0.397	$\pm$	0.067	&	0.226	$\pm$	0.054	&	0.369	$\pm$	0.086	&	0.201	$\pm$	0.089	&	12.30	$\pm$	0.09\\
J105640+013941 &	0.348	&	1.318	$\pm$	0.213	&	0.749	$\pm$	0.185	&	2.040	$\pm$	0.433	&	1.710	$\pm$	0.436	&	$\geq$	13.02\\	
J130811+113609 &	0.349	&	1.068	$\pm$	0.121	&	0.815	$\pm$	0.119	&	1.646	$\pm$	0.157	&	1.734	$\pm$	0.195	&	$\geq$	12.97\\	
J161018+042631 &	0.363	&	0.293	$\pm$	0.054	&	0.225	$\pm$	0.055	&	0.391	$\pm$	0.153	&	0.365	$\pm$	0.092	&	$\geq$	12.27\\	
J162957+423051 &	0.378	&	0.734	$\pm$	0.146	&	0.498	$\pm$	0.148	&	0.885	$\pm$	0.224	&	0.942	$\pm$	0.207	&	$\geq$	12.89\\	
J081336+481302 &	0.437	&	0.619	$\pm$	0.042	&	0.290	$\pm$	0.047	&	2.073	$\pm$	0.359	&	0.515	$\pm$	0.252	&	$\geq$	12.87\\	
J212727+082724 &	0.439	&	0.535	$\pm$	0.104	&	0.456	$\pm$	0.089	&	0.238	$\pm$	0.096	&	0.080	$\pm$	0.091	&	$\geq$	12.31\\	
J104923+012224 &	0.472	&	0.582	$\pm$	0.054	&	0.236	$\pm$	0.057	&	3.322	$\pm$	0.462	&	2.644	$\pm$	0.439	&	$\geq$	13.19\\	
J143614+105905 &	0.478	&	0.833	$\pm$	0.121	&	0.551	$\pm$	0.107	&	0.752	$\pm$	0.135	&	0.448	$\pm$	0.134	&	12.64	$\pm$	0.06\\
J015701+135503 &	0.484	&	1.760	$\pm$	0.184	&	1.599	$\pm$	0.318	&	2.652	$\pm$	0.322	&	2.838	$\pm$	0.277	&	$\geq$	13.37\\	
J125244+642103 &	0.512	&	1.099	$\pm$	0.096	&	0.696	$\pm$	0.076	&	1.518	$\pm$	0.824	&	2.189	$\pm$	1.119	&	$\geq$	12.99\\	
J083553+154139 &	0.531	&	1.064	$\pm$	0.106	&	0.726	$\pm$	0.117	&	3.377	$\pm$	0.911	&	1.320	$\pm$	0.962	&	13.08	$\pm$	0.12\\
J132803+352152 &	0.532	&	0.679	$\pm$	0.095	&	0.439	$\pm$	0.096	&	0.186	$\pm$	0.098	&	0.244	$\pm$	0.110	&	$\geq$	12.30\\	
J074816+422509 &	0.558	&	0.314	$\pm$	0.036	&	0.148	$\pm$	0.030	&	0.290	$\pm$	0.104	&	0.321	$\pm$	0.122	&	$\geq$	12.44\\	
J004800+022514 &	0.598	&	0.594	$\pm$	0.101	&	0.297	$\pm$	0.096	&	1.324	$\pm$	0.282	&	0.560	$\pm$	0.328	&	12.78	$\pm$	0.10\\
J132657+405018 &	0.611	&	0.723	$\pm$	0.080	&	0.548	$\pm$	0.078	&	1.149	$\pm$	0.152	&	0.683	$\pm$	0.147	&	$\geq$	12.82\\	
J160343+244836 &	0.656	&	0.723	$\pm$	0.080	&	0.548	$\pm$	0.078	&	0.957	$\pm$	0.277	&	0.612	$\pm$	0.259	&	12.76	$\pm$	0.09\\
%	1353	&	53083	&	569	&	0.511	&	0.715	$\pm$	0.143	&	0.44	$\pm$	0.118	&	1.098	$\pm$	1.072	&	1.374	$\pm$	1.354	&	$\geq$	12.81
%	724	&	52254	&	613	&	0.53	&	0.697	$\pm$	0.137	&	0.488	$\pm$	0.11	&	1.666	$\pm$	1.036	&	1.938	$\pm$	0.936	&	$\geq$	12.97
\hline

\end{tabular}
\label{TableSix}
\end{table*}

%1417	53141	52           J155752+342140
%1582	52939	481   J075031+192754
%1740	53050	438   J091958+111152
%2575	54085	81   J085917+105509
%1620	53137	499   J114339+073105
%4031	55604	14   J142536-001702
%0448 	51900	466   J085045+563618
%1267	52932	216   J082312+264415
%1415	52885	501   J165743+221149
%2613	54481	611   J124300+204246
%1785	54439	273   J085010+593118
%3832	55289	162   J102935-012138
% XXXX  1946	53432	501   J094927+314110 Too large Na?? 
%0794  	52376	568   J152800+535223
%1171	52753	416   J161649+415416
%4732	55648	257   J105640+013941
%1696	53116	107   J130811+113609
%4807	55687	732   J161018+042631
%0816	52379	517   J162957+423051
%3693	55208	424   J081336+481302
%4089	55470	402   J212727+082724
%4733	55649	131   J104923+012224
%1711	53535	339   J143614+105905
%4657	55591	159   J015701+135503
%0601	52316	110  J125244+642103
%2276	53712	54  J083553+154139
%3983	55603	252  J132803+352152
%3669	55481	92  J074816+422509
%4306	55584	686  J004800+022514
%4707	55653	894  J132657+405018
%3934	55336	662  J160343+244836

%% file: TableOne.tex
%%\bgroup
\def\arraystretch{1.2}
\begin{table*}
%%\tiny
%\caption{The median stacking measurements of the $435$ \CaII~absorbers in the Sardane et al. (2014) survey.}
\caption{Rest equivalent width (REW, W$_0$) measurements off five different normalized composite 
spectra, including the composite spectrum formed using the full sample of $435$ \CaII~absorber spectra. 
The rationale behind forming four additional composite spectra by dividing the full sample into subsamples is explained in the text.}

\begin{tabular}{l@{\extracolsep{1.0cm}}c@{\extracolsep{0.5cm}}c@{\extracolsep{1.0cm}}c@{\extracolsep{0.1cm}}c@{\extracolsep{1.0cm}}c}
 \toprule
~~~Line~~    &	\multicolumn{5}{c}{REW	($\mathrm{\AA}$)}    \\
\cline{2-6}

~~\nodata~~ &  \multicolumn{1}{c}{Full Sample} &  \multicolumn{1}{c}{$\mathrm{W_0^{\lambda3934} < 0.7\AA}$} &  
\multicolumn{1}{c}{$\mathrm{W_0^{\lambda3934} \geq 0.7\AA}$} &   
 $\mathrm{W_0^{\lambda2796}/W_0^{\lambda3934} < 1.8}$ &  \multicolumn{1}{c}{$\mathrm{W_0^{\lambda2796}/W_0^{\lambda3934} \geq 1.8}$} \\
\midrule
%\NiII~1703$^\star$	&	0.131	$\pm$	0.039	& 	0.079	$\pm$ 	0.019 &	\nodata &	0.132	$\pm$	0.032	&	0.168	$\pm$	0.044 	\\
%\NiII~1709$^\star$	&	0.081	$\pm$	0.038	&  	$\leq$ 0.064	  &	\nodata &	0.078	$\pm$	0.032	&	0.094	$\pm$	0.030 	 		\\
%%\SI~1807	&	$\leq$ 0.076 		&	$\leq$ 0.086 	  &	\nodata &	0.079	$\pm$	0.023	&	$\leq$ 0.156		\\
%%\SiI~1845	&	0.064	$\pm$	0.027	&	$\leq$ 0.154 	  & $\leq$ 0.460~ &	0.108	$\pm$	0.047	&	0.065	$\pm$	0.030	\\

\NiII~1741	&	0.133	$\pm$	0.039   &	0.107	$\pm$	0.029	&	0.054	$\pm$	0.022 &	\nodata &	0.130	$\pm$	0.029		\\
\NiII~1751	&	0.081	$\pm$	0.031   &	0.060	$\pm$	0.026	&	0.034	$\pm$	0.018 &	\nodata &	0.059	$\pm$	0.018		\\
\SiII~1808	&	0.155	$\pm$	0.029	&	0.160   $\pm$	0.024	&	0.151	$\pm$	0.028 & $\leq$ 0.437~ &	0.159	$\pm$	0.025		\\
\AlIII~1854	&	0.424	$\pm$	0.112	&	0.378	$\pm$	0.078	&	0.525 $\pm$	0.086 	  & $\leq$ 0.348~ &	0.512	$\pm$	0.066		\\
\AlIII~1862	&	0.286	$\pm$	0.106	&	0.322	$\pm$	0.052	&	0.363 $\pm$	0.071 	  & $\leq$ 0.399~ &	0.315	$\pm$	0.044		\\
\FeII~1901	&	$\leq$ 0.050		&	$\leq$ 0.032		&	$\leq$ 0.068 	  & $\leq$ 0.322~ &  $\leq$	    0.055		\\
\TiII~1910	&	$\leq$ 0.058		&	0.084 $\pm$ 0.021 		&	$\leq$ 0.070 	  & $\leq$ 0.223~ &  0.067$\pm$	    0.020		\\
\CoII~1941\textsuperscript{*} &	0.094	$\pm$	0.020	&	0.063	$\pm$	0.022	&	$\leq$ 0.064 	  & $\leq$ 0.323~ &	0.083	$\pm$	0.024   	\\
\CoII~2012\textsuperscript{*} &	$\leq$ 0.058		&	$\leq$ 0.028 		&	$\leq$ 0.089 	  & $\leq$ 0.344~ &	0.066	$\pm$	0.024    	\\
\ZnII~2026	&   0.114	$\pm$	0.020	    &	0.070	$\pm$	0.015   &	0.103 $\pm$	0.035 	  & $\leq$ 0.205\textsuperscript{\textdagger} &	0.079	$\pm$	0.031 \\
\CrII~2026	&	0.004 $\pm$ 0.002		&	0.003 $\pm$ 0.001 		&	0.004 $\pm$ 0.001 	  &	$\leq$ 0.005~ &	0.005	$\pm$	0.002   	\\
\MgI~~2026	&	0.023 $\pm$ 0.001		&	0.019 $\pm$	0.001 	    &	0.029 $\pm$	0.001 	  &	$\leq$ 0.015~ &	0.026	$\pm$	0.001   	\\
\CrII~2056	&	0.084 $\pm$	0.023		&	0.082	$\pm$	0.010	&	0.085 $\pm$	0.021	  & $\leq$ 0.303~ &	0.100	$\pm$	0.036       \\
\CrII~2062	&   0.056 $\pm$	0.030		&	0.063	$\pm$	0.012	&	$\leq$	0.074     & $\leq$ 0.309~ &	0.075	$\pm$	0.027       \\
\ZnII~2062	&	0.069 $\pm$ 0.035		&	0.058	$\pm$	0.020	&	0.157 $\pm$	0.042 	  &	$\leq$ 0.146\textsuperscript{\textdaggerdbl} &	$\leq$ 0.088        \\
\CrII~2066	&	0.028	$\pm$	0.020	&	0.044	$\pm$	0.011	&	$\leq$ 0.064	  & $\leq$ 0.181~ &	0.048	$\pm$	0.024       \\
\CdII~2145	&	$\leq$ 0.034		&	$\leq$ 0.034 		&	$\leq$ 0.038      & $\leq$ 0.158~ &	$\leq$ 0.060            \\
\FeI~~2167	&	$\leq$ 0.054		&	$\leq$ 0.038 		&	$\leq$ 0.064 	  & $\leq$ 0.125~ &	$\leq$ 0.066            \\
\FeII~2249	&	0.126	$\pm$	0.026	&	0.090	$\pm$	0.019	&	0.108 $\pm$	0.030 	  &	0.097 $\pm$ 0.023~ &	0.130	$\pm$	0.027   \\
\FeII~2260	&	0.115	$\pm$	0.020	&	0.110	$\pm$	0.020	&	0.096 $\pm$	0.030  	  &	0.099 $\pm$ 0.036~ &	0.136	$\pm$	0.025   \\
\FeII~2344	&	1.140	$\pm$	0.027	&	0.976	$\pm$	0.015	&	1.260 $\pm$	0.039 	  & 0.391 $\pm$ 0.034~ &	1.269	$\pm$	0.029   \\	
\FeII~2374	&	0.751	$\pm$	0.024	&	0.618	$\pm$	0.021	&	0.819 $\pm$	0.033 	  &	0.148 $\pm$ 0.042~ &	0.840	$\pm$	0.026   \\	
\FeII~2382	&	1.398	$\pm$	0.025	&	1.279	$\pm$	0.019	&	1.475 $\pm$	0.032 	  &	0.486 $\pm$ 0.043~ &	1.635	$\pm$	0.029   \\
\FeI~~2463	&	$\leq$ 0.050		&	$\leq$ 0.034		&	$\leq$ 0.072	  & $\leq$ 0.138 & $\leq$ 0.054             \\
\FeI~~2484	&	$\leq$ 0.052		&	$\leq$ 0.036		&	$\leq$ 0.070 	  & $\leq$ 0.117 &	$\leq$ 0.056            \\
\FeI~2501	&	$\leq$ 0.050		&	$\leq$ 0.038		&	$\leq$ 0.068 	  & $\leq$ 0.098 &	$\leq$ 0.056            \\
\SiI~2515	&	$\leq$ 0.050		&	$\leq$ 0.038		&	$\leq$ 0.058 	  & $\leq$ 0.123 & $\leq$ 0.054             \\
\FeI~~2523	&	$\leq$ 0.046		&	$\leq$ 0.032		&	$\leq$ 0.056 	  & $\leq$ 0.110 &	$\leq$ 0.042            \\
\MnII~2576	&	0.226	$\pm$	0.029	&	0.183	$\pm$	0.018	&	0.299 $\pm$	0.030 	  & 0.236	$\pm$ 0.067 &	0.237	$\pm$	0.023   \\
\FeII~2586	&	1.115	$\pm$	0.029	&	0.954	$\pm$	0.021	&	1.247 $\pm$	0.033 	  & 0.711	$\pm$ 0.047 &	1.239	$\pm$	0.025   \\
\MnII~2594	&	0.164	$\pm$	0.026	&	0.137	$\pm$	0.017	&	0.197 $\pm$	0.024 	  & 0.181	$\pm$ 0.063 &	0.171	$\pm$	0.024   \\
\FeII~2600	&	1.472	$\pm$	0.029	&	1.321	$\pm$	0.020	&	1.616 $\pm$	0.034 	  & 0.770	$\pm$ 0.039 &	1.641	$\pm$	0.024   \\
\MnII~2606	&	0.104	$\pm$	0.023	&	0.100	$\pm$	0.017	&	0.115 $\pm$	0.020 	  & 0.109   $\pm$ 0.038 &	0.123	$\pm$	0.028       \\
\MgII~2796	&	1.940	$\pm$	0.024	&	1.785	$\pm$	0.021	&	2.068 $\pm$	0.032 	  & 1.235	$\pm$ 0.038	 &	2.283	$\pm$	0.024   \\
\MgII~2803	&	1.803	$\pm$	0.023	&	1.595	$\pm$	0.020	&	2.054 $\pm$	0.032 	  & 1.175	$\pm$ 0.033	 &	2.065	$\pm$	0.022   \\
\MgI~2852	&	0.742	$\pm$	0.028	&	0.614	$\pm$	0.020	&	0.919 $\pm$	0.037 	  & 0.474	$\pm$ 0.032	 &	0.818	$\pm$	0.021   \\
\FeI~2967	&	$\leq$ 0.046	  	&	$\leq$ 0.032 		&	$\leq$ 0.060 	  & $\leq$ 0.074 &  $\leq$ 0.046            \\
\FeI~3021	&	$\leq$ 0.046	  	&	$\leq$ 0.038 		&	$\leq$ 0.056 	  & $\leq$ 0.069 &	$\leq$ 0.046            \\
\TiII~3073	&	0.039	$\pm$	0.014	&	0.039	$\pm$	0.012	&	0.036 $\pm$ 0.018 	  & 0.057 $\pm$ 0.018 &	0.032	$\pm$	0.012       \\
\TiII~3230	&	0.035	$\pm$	0.014	&	0.035	$\pm$	0.014	&	0.048 $\pm$ 0.020 	  & 0.045 $\pm$ 0.019 &  0.030   $\pm$   0.011 		\\
\TiII~3242	&	0.046	$\pm$	0.015	&	0.050	$\pm$	0.015	&	0.052 $\pm$	0.019 	  & 0.068 $\pm$ 0.020 &	0.052	$\pm$	0.012       \\
\TiII~3384	&	0.082	$\pm$	0.016	&	0.091	$\pm$	0.016	&	0.089 $\pm$	0.020 	  & 0.111 $\pm$ 0.017 &	0.079	$\pm$	0.013       \\
\FeI~3720	&	$\leq$ 0.042		&		$\leq$ 0.030 	&	$\leq$ 0.056 	  & $\leq$ 0.079 &	$\leq$ 0.044            \\
\CaII~3934	&	0.703	$\pm$	0.021	&	0.493	$\pm$	0.014	&	1.012 $\pm$	0.032 	  & 0.885 $\pm$ 0.026 &	0.636	$\pm$	0.017       \\
\CaII~3969	&	0.418	$\pm$	0.023	&	0.312	$\pm$	0.016	&	0.572 $\pm$	0.032 	  & 0.470 $\pm$ 0.026 &	0.378	$\pm$	0.019       \\
\CaI~4227	&	$\leq$ 0.050		&		$\leq$ 0.026 	&	$\leq$ 0.066 	  & $\leq$ 0.071 &	$\leq$ 0.032            \\
\NaI~5891	&	0.118	$\pm$	0.041	&	0.079	$\pm$	0.026	&	0.194 $\pm$	0.067 	  & 0.235 $\pm$ 0.054 &	0.269	$\pm$	0.048       \\
\NaI~5897	&	0.089	$\pm$	0.040	&	0.074	$\pm$	0.025	&	0.182 $\pm$	0.075 	  & $\leq$ 0.104 &	0.174	$\pm$	0.044       \\
\bottomrule
%\CoII~1941	&	0.055	$\pm$	0.021	&	0.063	$\pm$	0.022	&	$\leq$ 0.064 	  & $\leq$ 0.323~ &	0.082	$\pm$	0.033   	\\
%\TiII~1910b	&	$\leq$ 0.066		&	$\leq$ 0.066 		&	$\leq$ 0.084 	  & $\leq$ 0.229~ &	0.067	$\pm$	0.031   	\\

\end{tabular}
\label{TableOne}
  \begin{tablenotes}[flushleft]
    \scriptsize
    \item \textsuperscript{*} Tentative detections
    \item \textsuperscript{\textdagger } Upper limit inferred using the $2\sigma$ upper limit for \CrII+\MgI+\ZnII.
    \item \textsuperscript{\textdaggerdbl} Upper limit inferred using the $2\sigma$ upper limit for \CrII+\ZnII.    
  \end{tablenotes}
\end{table*}

%% file: TableTwo.tex
%%%%%%%%%%%%%%%%%%%%%%%%%%%%%%%%%%%%%%%%%%%%%%%%%%%%%%%%%%%%%%%%%%%%%%%%%%%%%%%%%%%%%%%%%%%%%%%%%%%%%%%%%%%%%%%%%%%%%%%%%%%%%%%%%%%%%%%%%%%%%%%%%%%%%%%%%%%%%%%%%%%%%%%%%%%%%%%%%%%%
%%%%%%%%%%%%%%%%%%%%%%%%%%%%%%%%%%%%%%%%%%%%%%%%%%%%%%%%%%%%%%%%%%%%%%%%%%%%%%%%%%%%%%%%%%%%%%%%%%%%%%%%%%%%%%%%%%%%%%%%%%%%%%%%%%%%%%%%%%%%%%%%%%%%%%%%%%%%%%%%%%%%%%%%%%%%%%%%%%%%
\bgroup
\def\arraystretch{1.5}

\begin{table*}
%\small
%\caption{The column densities inferred from the median REWs of the stacks.}
\caption{Metal-line column densities derived from the normalized composite spectra.}
\begin{tabular}{lcccccc}
\toprule
	&		&	& $\mathrm{log~N}$ [$\mathrm{atoms~per~cm^2}$]	 	   \\  
\cline{2-6}
Ion & \multicolumn{1}{c|}{Full Sample} & \multicolumn{1}{c|}{$W_0^{\lambda3934} < 0.7\mathrm{\AA}$} & \multicolumn{1}{c|}	{$W_0^{\lambda3934} \geq 0.7\mathrm{\AA}$}	
& {$W_0^{\lambda2796}/W_0^{\lambda3934} < 1.8$} & {$W_0^{\lambda2796}/W_0^{\lambda3934} \geq 1.8$} \\

\midrule
Si$^{+1}$  & 15.39	$\pm$	0.08 	&  15.40 $\pm$	0.07 & 15.38 $\pm$ 0.08 & \nodata          & 15.40	$\pm$	0.07 \\
Al$^{+2}$  & 13.44	$\pm$	0.09 	&  13.36 $\pm$	0.09 & 13.51 $\pm$ 0.07 & $\leq$ 13.54 & 13.49	$\pm$	0.06 \\
Zn$^{+1}$  & 12.81	$\pm$	0.07 	&  12.63 $\pm$	0.08 & 12.88 $\pm$ 0.10 & $\leq$ 13.35 & 12.69	$\pm$	0.15 \\
Cr$^{+1}$  & 13.31	$\pm$	0.09 	&  13.33 $\pm$	0.04 & 13.08 $\pm$ 0.12 & $\leq$ 14.23 & 13.41	$\pm$	0.08 \\
Fe$^{+1}$  & 15.12	$\pm$	0.06 	&  15.01 $\pm$	0.06 & 15.05 $\pm$ 0.09 & 15.03 $\pm$ 0.09 & 15.13	$\pm$	0.06 \\
Mn$^{+1}$  & 13.01	$\pm$	0.04 	&  12.94 $\pm$	0.03 & 13.00 $\pm$ 0.04 & 12.82 $\pm$ 0.10 & 13.05	$\pm$	0.03 \\
Ti$^{+1}$  & 12.42	$\pm$	0.06    &  12.46 $\pm$	0.06 & 12.41 $\pm$ 0.08 & 12.55 $\pm$ 0.05 & 12.41	$\pm$	0.05 \\
Fe$^{~0}$   &    $\leq$ 12.99   &  $\leq$ 13.11 & $\leq$ 12.84 & $\leq$ 13.04 &   $\leq$ 12.99 \\
Ca$^{+1}$   & $\geq$ 12.97        & $\geq$12.84 & $\geq$13.11  & 13.02 $\pm$ 0.02 & $\geq$12.87 \\  % % $\pm$	0.02 
Na$^{~0}$   & $\geq$ 11.83        & $\geq$11.68 & $\geq$12.14  & $\geq$12.06	  & $\geq$12.17\\
\bottomrule
\end{tabular}
\label{TableTwo}
\end{table*}

%%%%%%%%%%%%%%%%%%%%%%%%%%%%%%%%%%%%%%%%%%%%%%%%%%%%%%%%%%%%%%%%%%%%%%%%%%%%%%%%%%%%%%%%%%%%%%%%%%%%%%%%%%%%%%%%%%%%%%%%%%%%%%%%%%%%%%%%%%%%%%%%%%%%%%%%%%%%%%%%%%%%%%%%%%%%%%%%%%%%
%%%%%%%%%%%%%%%%%%%%%%%%%%%%%%%%%%%%%%%%%%%%%%%%%%%%%%%%%%%%%%%%%%%%%%%%%%%%%%%%%%%%%%%%%%%%%%%%%%%%%%%%%%%%%%%%%%%%%%%%%%%%%%%%%%%%%%%%%%%%%%%%%%%%%%%%%%%%%%%%%%%%%%%%%%%%%%%%%%%%

%FeI 2sigma Upper Limits:

%W3934 <  0.7A       <= 1.3E+13/cm^3
%W3934 >= 0.7A       <= 7.0E+12/cm^3
%W2796/W3934 <   1.8 <= 1.1E+13/cm^3
%W2796/W3934 >=  1.8 <= 9.7E+12/cm^3
%FULL SAMPLE         <= 9.8E+12/cm^3

%% file: TableThree.tex
\bgroup
\def\arraystretch{1.5}

\begin{table*}
%\small
%\caption{Elemental abundances relative to Zn for the various sub-populations of \CaII~absorbers. The measurements are
%derived from the REWs of the median stack. Abundance ratios relative to Zn are used as an 
%indicator in assessing the dust depletion of absorbers. Solar abundances are taken from Asplund et al. (2009).}
\caption{Elemental abundances relative to Zn for the various (sub)samples of \CaII\ absorbers. The determinations are derived using the methods discussed in the text.}

\begin{tabular}{lcccccc}
\toprule			&	& & $\mathrm{[X/Zn]}$ & & \\  
\cline{2-6}
X & \multicolumn{1}{c|}{Full Sample} & \multicolumn{1}{c|}{$W_0^{\lambda3934} < 0.7\mathrm{\AA}$} & \multicolumn{1}{c|}{$W_0^{\lambda3934} \geq 0.7\mathrm{\AA}$}	
& {$W_0^{\lambda2796}/W_0^{\lambda3934} < 1.8$} & {$W_0^{\lambda2796}/W_0^{\lambda3934} \geq 1.8$}  \\

\midrule
Cr  & $-0.58 \pm 0.12$ &  $-0.38 \pm 0.09 $	& $ -0.88 \pm 0.16 $ &  \nodata         &$-0.36  \pm 0.18 $ 	\\
Si  & $-0.37 \pm 0.11$ &  $-0.17 \pm 0.10 $	& $ -0.45 \pm 0.13 $ &  \nodata         &$-0.23  \pm 0.10 $ 	\\
Mn  & $-0.67 \pm 0.11$ &  $-0.56 \pm 0.11 $	& $ -0.75 \pm 0.11 $ &  $\geq -1.49$ &$-0.51  \pm 0.13 $ 	\\
Ti  & $-0.78 \pm 0.10$ &  $-0.55 \pm 0.12 $	& $ -0.86 \pm 0.13 $ &  $\geq -1.19$ &$-0.66  \pm 0.16 $     \\
Fe  & $-0.63 \pm 0.09$ &  $-0.55 \pm 0.09 $	& $ -0.77 \pm 0.09 $ &  $\geq -1.26$ &$-0.50  \pm 0.17 $ 	\\
%Co  & $+0.37 \pm 0.20$ &  $+0.29 \pm 0.23 $	& $ +0.25 \pm 0.29 $ &  \nodata         &$+0.66  \pm 0.19 $ 	\\
Ni  & $-0.42 \pm 0.16$ &  $-0.34 \pm 0.16 $	& $ -0.87 \pm 0.19 $ &	\nodata         &$-0.37  \pm 0.14  $     \\

\bottomrule
\end{tabular}
\label{TableThree}
\end{table*}

%%%%%%%%%%%%%%%%%%%%%%%%%%%%%%%%%%%%%%%%%%%%%%%%%%%%%%%%%%%%%%%%%%%%%%%%%%%%%%%%%%%%%%%%%%%%%%%%%%%%%%%%%%%%%%%%%%%%%%%%%%%%%%%%%%%%%%%%%%%%%%%%%%%%%%%%%%%%%%%%%%%%%%%%%%%%%%%%%%%%
%%%%%%%%%%%%%%%%%%%%%%%%%%%%%%%%%%%%%%%%%%%%%%%%%%%%%%%%%%%%%%%%%%%%%%%%%%%%%%%%%%%%%%%%%%%%%%%%%%%%%%%%%%%%%%%%%%%%%%%%%%%%%%%%%%%%%%%%%%%%%%%%%%%%%%%%%%%%%%%%%%%%%%%%%%%%%%%%%%%%

%Na  &  $\geq -2.88  $    & $ \geq -2.50$ &  $\geq -2.50$ &$\geq-2.39$ & $\geq-2.73$  \\
%Na  &  $-2.88  $    & $ -2.50 \pm 0.16 $ &  $-2.50\pm 0.14$ &$-2.39  \pm 0.12 $ & $-2.73 \pm 0.16$  \\
%\textcolor{red}{Ca}  &  $\geq -1.57 $	&  $ \geq -1.46 $  & $\geq  -2.11$ & $\geq -1.13 $  & $\geq -1.47 $ \\

%% file: TableFour.tex
%%%%%%%%%%%%%%%%%%%%%%%%%%%%%%%%%%%%%%%%%%%%%%%%%%%%%%%%%%%%%%%%%%%%%%%%%%%%%%%%%%%%%%%%%%%%%%%%%%%%%%%%%%%%%%%%%%%%%%%%%%%%%%%%%%%%%%%%%%%%%%%%%%%%%%%%%%%%%%%%%%%%%%%%%%%%%%%%%%%%
%%%%%%%%%%%%%%%%%%%%%%%%%%%%%%%%%%%%%%%%%%%%%%%%%%%%%%%%%%%%%%%%%%%%%%%%%%%%%%%%%%%%%%%%%%%%%%%%%%%%%%%%%%%%%%%%%%%%%%%%%%%%%%%%%%%%%%%%%%%%%%%%%%%%%%%%%%%%%%%%%%%%%%%%%%%%%%%%%%%%

\bgroup
\def\arraystretch{1.5}
\begin{table*}
%\small
%\caption{Elemental abundances relative to Fe for the various sub-populations of \CaII~absorbers. The measurements are
%derived from the REWs of the median stack. Abundance ratios relative to Zn are used as an 
%indicator in assessing the dust depletion of absorbers. Solar abundances are taken from Asplund et al. (2009). 
%The errors are propagated assuming Gaussian error distributions.}
\caption{Elemental abundances relative to Fe for the various (sub)samples of \CaII\ absorbers. The determinations are derived using the methods discussed in the text.}

\begin{tabular}{ccccccc}
\toprule			&	& & $\mathrm{[X/Fe]}$ & & \\  
\cline{2-6}
X & \multicolumn{1}{c|}{Full Sample} &  \multicolumn{1}{c|}{$W_0^{\lambda3934} < 0.7\mathrm{\AA}$} & \multicolumn{1}{c|}{$W_0^{\lambda3934} \geq 0.7\mathrm{\AA}$}	
& {$W_0^{\lambda2796}/W_0^{\lambda3934} < 1.8$} & {$W_0^{\lambda2796}/W_0^{\lambda3934} \geq 1.8$}  \\

\midrule
Cr  & $ +0.05 \pm 0.11$	  &  $+0.17 \pm 0.07 $ & $-0.12 \pm 0.15$  &  $\leq+1.06$  & $+0.14 \pm 0.10$ \\
Si  & $ +0.26 \pm 0.10$   &  $+0.38 \pm 0.09 $ & $+0.31 \pm 0.12$  &  \nodata         & $+0.26 \pm 0.09$ \\
Mn  & $ -0.04 \pm 0.07$   &  $-0.01 \pm 0.07 $ & $+0.01 \pm 0.10$  &  $0.14 \pm 0.13$ & $-0.01 \pm 0.07$ \\ 
Ti  & $ -0.15 \pm 0.09$   &  $~~0.00 \pm 0.08$ & $-0.09 \pm 0.12$  &  $0.07 \pm 0.10$ & $-0.16 \pm 0.08$ \\ 
Zn  & $ +0.63 \pm 0.09$   &  $+0.55 \pm 0.09 $ & $+0.77 \pm 0.09 $ &  $\leq 1.26$ & $+0.50 \pm 0.17$ \\
Ni  & $ +0.21 \pm 0.12$   &  $+0.21 \pm 0.12 $ & $-0.11 \pm 0.17$  &  \nodata         & $+0.13 \pm 0.10$ \\
%Co  & $ +1.00 \pm 0.16$	  &  $+0.84 \pm 0.19 $ & $+1.01 \pm 0.27$  &  \nodata         & $+1.16 \pm 0.15$ \\

\bottomrule
\noindent
\vspace{-9ex}

\end{tabular}
%\begin{tablenotes}[flushleft]
%    \item \textsuperscript{\textdagger } 
%\end{tablenotes}
\label{TableFour}
\end{table*}

%%%%%%%%%%%%%%%%%%%%%%%%%%%%%%%%%%%%%%%%%%%%%%%%%%%%%%%%%%%%%%%%%%%%%%%%%%%%%%%%%%%%%%%%%%%%%%%%%%%%%%%%%%%%%%%%%%%%%%%%%%%%%%%%%%%%%%%%%%%%%%%%%%%%%%%%%%%%%%%%%%%%%%%%%%%%%%%%%%%%
%%%%%%%%%%%%%%%%%%%%%%%%%%%%%%%%%%%%%%%%%%%%%%%%%%%%%%%%%%%%%%%%%%%%%%%%%%%%%%%%%%%%%%%%%%%%%%%%%%%%%%%%%%%%%%%%%%%%%%%%%%%%%%%%%%%%%%%%%%%%%%%%%%%%%%%%%%%%%%%%%%%%%%%%%%%%%%%%%%%%
%S   &  $-1.32 \pm 0.16 $ & $-1.46 \pm 0.17$  &  \nodata         & $-1.37 \pm 0.12$ & $ -1.31 \pm 0.17$	 \\
%Al\tablenotemark{a}  &  $ \geq -0.60 $ & $-0.50  \pm 0.11$  & $\leq -0.43$  & $ -0.59 \pm 0.08$  & $ -0.63  \pm 0.11$   \\
%\tablenotetext{a}{Inferred from the weighted average of \AlIII $\lambda$~1854 and \AlIII 
%$\lambda$~1862 profiles, which are found to be saturated.}

%% file: TableZero.tex
 \setlength{\tabcolsep}{0.0em}
\begin{table*}
\centering
	\addtocounter{table}{0} \small
%\caption{The reddening measurements results: LMC, SMC and the MW dust models for populations of \CaII.}
\caption{Results on extinction and reddening for the full sample and four subsamples of \CaII\ absorbers. 
Parameters are given for the best-fit LMC, SMC, and MW extinction laws applied to the fluxed \CaII\ absorber 
composite spectra, relative to the unabsorbed reference composite spectra, at $\lambda_{rest} \geq 2500$ \AA. 
The parameter $\mathcal{R}$ is the absorbed-to-unabsorbed flux ratio at 2200 \AA. For the \CaII\ absorber 
(sub)samples $\mathcal{R}$ is observed. For the best-fit extinction models $\mathcal{R}$ is predicted.}

\small
\begin{tabular}{l@{\extracolsep{0.2cm}}c@{\extracolsep{0.2cm}}c@{\extracolsep{0.2cm}}c@{\extracolsep{0.2cm}}c@{\extracolsep{0.2cm}}c@{\extracolsep{0.2cm}}c@{\extracolsep{0.2cm}}c@{\extracolsep{0.2cm}}c@{\extracolsep{0.2cm}}c@{\extracolsep{0.2cm}}c}
 \toprule
&\multicolumn{2}{c|}{ Full Sample} & \multicolumn{2}{c|}{ $W_0^{\lambda3934} < 0.7\mathrm{\AA}$}
&\multicolumn{2}{c|}{ $W_0^{\lambda3934} \geq 0.7\mathrm{\AA}$ } &\multicolumn{2}{c|}{ $\mathrm{W_0^{\lambda2796}/W_0^{\lambda3934} < 1.8}$}
&\multicolumn{2}{c|}{$\mathrm{W_0^{\lambda2796}/W_0^{\lambda3934} \geq 1.8}$}\\
\cmidrule{2-3}\cmidrule{4-5}\cmidrule{6-7}\cmidrule{8-9}\cmidrule{10-11}

Dust Law	&	$E(B-V)$ &	 $\mathcal{R}$	&	$E(B-V)$ & 	 	 $\mathcal{R}$	&
$E(B-V)$ &	 	 $\mathcal{R}$	 & $E(B-V)$ &		 $\mathcal{R}$		& 
$E(B-V)$ &	 	 $\mathcal{R}$		\\
\nodata	&	[mag] &	 \nodata	&	[mag] &\nodata&
[mag] &	\nodata	& [mag] &	 \nodata	& 
[mag] &	 \nodata	\\

\midrule
LMC			  &	0.033  & 0.81  &	0.012 & 0.92	  &	0.051 & 0.70 &	   0.030   & 0.79    &	0.022 &	0.85   \\
SMC			  &	0.026  & 0.82  &	0.010 & 0.92 	  &	0.043 & 0.72 &	   0.025   & 0.80    &	0.018 &	0.86   \\
MW			  &	0.035  & 0.77  &	0.011 & 0.89 	  &	0.047 & 0.65 &	   0.027   & 0.75    &	0.020 &	0.80   \\
\CaII\ Sample &  -     & 0.83  &    -     & 0.95      & -     & 0.73 &      -      & \nodata &    -   & 0.83   \\
 \bottomrule
\end{tabular}
\label{TableZero}
\end{table*}
%%%FULL TABLE
%% Weak ~ \pm 3-4%
%% Strong ~\pm \sim 1%
%% Full sample ~
%LMC	&	0.0100 $\pm$ 0.0003	&	0.079 $\pm$ 0.003	&	0.050	$\pm$ 0.0004	&	0.394 $\pm$ 0.003 &	0.0100 $\pm$ 0.0003	&	0.079 $\pm$ 0.003&	0.0100 $\pm$ 0.0003	&	0.079 $\pm$ 0.003&	0.0100 $\pm$ 0.0003	&	0.079 $\pm$ 0.003\\
%SMC	&	0.0080 $\pm$ 0.0003	&	0.075 $\pm$ 0.002	&	0.043	$\pm$ 0.0003	&	0.362 $\pm$ 0.003 &	0.0080 $\pm$ 0.0003	&	0.075 $\pm$ 0.002&	0.0080 $\pm$ 0.0003	&	0.075 $\pm$ 0.002&	0.0080 $\pm$ 0.0003	&	0.075 $\pm$ 0.002\\
%MW	&	0.0100 $\pm$ 0.0003	&	0.095 $\pm$ 0.003	&	0.047	$\pm$ 0.0003	&	0.448 $\pm$ 0.003 &	0.0100 $\pm$ 0.0003	&	0.095 $\pm$ 0.003&	0.0100 $\pm$ 0.0003	&	0.095 $\pm$ 0.003&	0.0100 $\pm$ 0.0003	&	0.095 $\pm$ 0.003\\
%LMC	&	0.012&0.094	    &	0.051&0.41 &	   0.030 &0.239 &	0.022 &	0.239 &	0.033  & 0.26  \\
%SMC	&	0.010&0.086 	&	0.043&0.36 &	   0.025 &0.213 &	0.018 &	0.213 &	0.026  & 0.22  \\
%MW	&	0.011&0.106 	&	0.047&0.44 &	   0.027 &0.255 &	0.020 &	0.188 &	0.035  & 0.32  \\

%%%%%%%%%%%%%%%%%%%%%%%%%%%%%%%%%%%%%%%%%%%%%%%%%%%%%%%%%%%%%%%%%%%%%%%%%%%%%%%%%%%%%%%%%%%%%%%%%%%%%%%%%%%%%%%%%%%%%%%%%%%%%%%%%%%%%%%%%%%%%%%%%%%%%%%%%%%%%%%%%%%%%%%%%%%%%%%%%%%%
%%%%%%%%%%%%%%%%%%%%%%%%%%%%%%%%%%%%%%%%%%%%%%%%%%%%%%%%%%%%%%%%%%%%%%%%%%%%%%%%%%%%%%%%%%%%%%%%%%%%%%%%%%%%%%%%%%%%%%%%%%%%%%%%%%%%%%%%%%%%%%%%%%%%%%%%%%%%%%%%%%%%%%%%%%%%%%%%%%%%

%% file: CaIIDepletionArxiV_GSardane.bbl
\begin{thebibliography}{79}
\expandafter\ifx\csname natexlab\endcsname\relax\def\natexlab#1{#1}\fi

\bibitem[{{Abazajian} {et~al}\mbox{.}(2009){Abazajian}, {Adelman-McCarthy},
  {Ag{\"u}eros}, {Allam}, {Allende Prieto}, {An}, {Anderson}, {Anderson},
  {Annis}, {Bahcall}, \& et~al.}]{2009ApJS..182..543A}
{Abazajian} K.~N. {et~al.}, 2009, ApJS, 182, 543

\bibitem[{{Ahn} {et~al}\mbox{.}(2012){Ahn}, {Alexandroff}, {Allende Prieto},
  {Anderson}, {Anderton}, {Andrews}, {Aubourg}, {Bailey}, {Balbinot}, {Barnes},
  \& et~al.}]{2012ApJS..203...21A}
{Ahn} C.~P. {et~al.}, 2012, ApJS, 203, 21

\bibitem[{{Akerman} {et~al}\mbox{.}(2005){Akerman}, {Ellison}, {Pettini}, \&
  {Steidel}}]{2005A&A...440..499A}
{Akerman} C.~J., {Ellison} S.~L., {Pettini} M., {Steidel} C.~C., 2005, A\&A,
  440, 499

\bibitem[{{Asplund} {et~al}\mbox{.}(2009){Asplund}, {Grevesse}, {Sauval}, \&
  {Scott}}]{2009ARA&A..47..481A}
{Asplund} M., {Grevesse} N., {Sauval} A.~J., {Scott} P., 2009, ARA\&A, 47, 481

\bibitem[{{Battisti} {et~al}\mbox{.}(2012){Battisti}, {Meiring}, {Tripp},
  {Prochaska}, {Werk}, {Jenkins}, {Lehner}, {Tumlinson}, \&
  {Thom}}]{2012ApJ...744...93B}
{Battisti} A.~J. {et~al.}, 2012, ApJ, 744, 93

\bibitem[{{Cooksey} {et~al}\mbox{.}(2013){Cooksey}, {Kao}, {Simcoe}, {O'Meara},
  \& {Prochaska}}]{2013ApJ...763...37C}
{Cooksey} K.~L., {Kao} M.~M., {Simcoe} R.~A., {O'Meara} J.~M., {Prochaska}
  J.~X., 2013, ApJ, 763, 37

\bibitem[{{Crawford}(1992)}]{1992MNRAS.259...47C}
{Crawford} I.~A., 1992, MNRAS, 259, 47

\bibitem[{{Crighton} {et~al}\mbox{.}(2013){Crighton}, {Bechtold}, {Carswell},
  {Dav{\'e}}, {Foltz}, {Jannuzi}, {Morris}, {O'Meara}, {Prochaska}, {Schaye},
  \& {Tejos}}]{2013MNRAS.433..178C}
{Crighton} N.~H.~M. {et~al.}, 2013, \mnras, 433, 178

\bibitem[{{Cui} {et~al}\mbox{.}(2005){Cui}, {Bechtold}, {Ge}, \&
  {Meyer}}]{2005ApJ...633..649C}
{Cui} J., {Bechtold} J., {Ge} J., {Meyer} D.~M., 2005, \apj, 633, 649

\bibitem[{{Dawson} {et~al}\mbox{.}(2013){Dawson}, {Schlegel}, {Ahn},
  {Anderson}, {Aubourg}, {Bailey}, {Barkhouser}, {Bautista}, {Beifiori},
  {Berlind}, {Bhardwaj}, {Bizyaev}, {Blake}, {Blanton}, {Blomqvist}, {Bolton},
  {Borde}, {Bovy}, {Brandt}, {Brewington}, {Brinkmann}, {Brown}, {Brownstein},
  {Bundy}, {Busca}, {Carithers}, {Carnero}, {Carr}, {Chen}, {Comparat},
  {Connolly}, {Cope}, {Croft}, {Cuesta}, {da Costa}, {Davenport}, {Delubac},
  {de Putter}, {Dhital}, {Ealet}, {Ebelke}, {Eisenstein}, {Escoffier}, {Fan},
  {Filiz Ak}, {Finley}, {Font-Ribera}, {G{\'e}nova-Santos}, {Gunn}, {Guo},
  {Haggard}, {Hall}, {Hamilton}, {Harris}, {Harris}, {Ho}, {Hogg}, {Holder},
  {Honscheid}, {Huehnerhoff}, {Jordan}, {Jordan}, {Kauffmann}, {Kazin},
  {Kirkby}, {Klaene}, {Kneib}, {Le Goff}, {Lee}, {Long}, {Loomis}, {Lundgren},
  {Lupton}, {Maia}, {Makler}, {Malanushenko}, {Malanushenko}, {Mandelbaum},
  {Manera}, {Maraston}, {Margala}, {Masters}, {McBride}, {McDonald}, {McGreer},
  {McMahon}, {Mena}, {Miralda-Escud{\'e}}, {Montero-Dorta}, {Montesano},
  {Muna}, {Myers}, {Naugle}, {Nichol}, {Noterdaeme}, {Nuza}, {Olmstead},
  {Oravetz}, {Oravetz}, {Owen}, {Padmanabhan}, {Palanque-Delabrouille}, {Pan},
  {Parejko}, {P{\^a}ris}, {Percival}, {P{\'e}rez-Fournon},
  {P{\'e}rez-R{\`a}fols}, {Petitjean}, {Pfaffenberger}, {Pforr}, {Pieri},
  {Prada}, {Price-Whelan}, {Raddick}, {Rebolo}, {Rich}, {Richards}, {Rockosi},
  {Roe}, {Ross}, {Ross}, {Rossi}, {Rubi{\~n}o-Martin}, {Samushia},
  {S{\'a}nchez}, {Sayres}, {Schmidt}, {Schneider}, {Sc{\'o}ccola}, {Seo},
  {Shelden}, {Sheldon}, {Shen}, {Shu}, {Slosar}, {Smee}, {Snedden}, {Stauffer},
  {Steele}, {Strauss}, {Streblyanska}, {Suzuki}, {Swanson}, {Tal}, {Tanaka},
  {Thomas}, {Tinker}, {Tojeiro}, {Tremonti}, {Vargas Maga{\~n}a}, {Verde},
  {Viel}, {Wake}, {Watson}, {Weaver}, {Weinberg}, {Weiner}, {West}, {White},
  {Wood-Vasey}, {Yeche}, {Zehavi}, {Zhao}, \& {Zheng}}]{2013AJ....145...10D}
{Dawson} K.~S. {et~al.}, 2013, AJ, 145, 10

\bibitem[{{Dessauges-Zavadsky}, {Prochaska} \&
  {D'Odorico}(2002){Dessauges-Zavadsky}, {Prochaska}, \&
  {D'Odorico}}]{2002A&A...391..801D}
{Dessauges-Zavadsky} M., {Prochaska} J.~X., {D'Odorico} S., 2002, A\&A, 391,
  801

\bibitem[{{Draine}(2011)}]{2011piim.book.....D}
{Draine} B.~T., 2011, {Physics of the Interstellar and Intergalactic Medium}

\bibitem[{{Edvardsson} {et~al}\mbox{.}(1995){Edvardsson}, {Pettersson},
  {Kharrazi}, \& {Westerlund}}]{1995A&A...293...75E}
{Edvardsson} B., {Pettersson} B., {Kharrazi} M., {Westerlund} B., 1995, A\&A,
  293, 75

\bibitem[{{Fitzpatrick}(1999)}]{1999PASP..111...63F}
{Fitzpatrick} E.~L., 1999, PASP, 111, 63

\bibitem[{{Foltz}, {Chaffee} \& {Black}(1988){Foltz}, {Chaffee}, \&
  {Black}}]{1988ApJ...324..267F}
{Foltz} C.~B., {Chaffee}, Jr. F.~H., {Black} J.~H., 1988, \apj, 324, 267

\bibitem[{{Fox} {et~al}\mbox{.}(2007){Fox}, {Petitjean}, {Ledoux}, \&
  {Srianand}}]{2007A&A...465..171F}
{Fox} A.~J., {Petitjean} P., {Ledoux} C., {Srianand} R., 2007, \aap, 465, 171

\bibitem[{{Fran{\c c}ois} {et~al}\mbox{.}(2004){Fran{\c c}ois}, {Matteucci},
  {Cayrel}, {Spite}, {Spite}, \& {Chiappini}}]{2004A&A...421..613F}
{Fran{\c c}ois} P., {Matteucci} F., {Cayrel} R., {Spite} M., {Spite} F.,
  {Chiappini} C., 2004, A\&A, 421, 613

\bibitem[{{Gordon} {et~al}\mbox{.}(2003){Gordon}, {Clayton}, {Misselt},
  {Landolt}, \& {Wolff}}]{2003ApJ...594..279G}
{Gordon} K.~D., {Clayton} G.~C., {Misselt} K.~A., {Landolt} A.~U., {Wolff}
  M.~J., 2003, ApJ, 594, 279

\bibitem[{{Herbert-Fort} {et~al}\mbox{.}(2006){Herbert-Fort}, {Prochaska},
  {Dessauges-Zavadsky}, {Ellison}, {Howk}, {Wolfe}, \&
  {Prochter}}]{2006PASP..118.1077H}
{Herbert-Fort} S., {Prochaska} J.~X., {Dessauges-Zavadsky} M., {Ellison} S.~L.,
  {Howk} J.~C., {Wolfe} A.~M., {Prochter} G.~E., 2006, PASP, 118, 1077

\bibitem[{{Kirkwood}(1979)}]{1979Biometrics.35}
{Kirkwood} T.~B., 1979, Biometrics, 35, 908

\bibitem[{{Kulkarni} {et~al}\mbox{.}(2005){Kulkarni}, {Fall}, {Lauroesch},
  {York}, {Welty}, {Khare}, \& {Truran}}]{2005ApJ...618...68K}
{Kulkarni} V.~P., {Fall} S.~M., {Lauroesch} J.~T., {York} D.~G., {Welty} D.~E.,
  {Khare} P., {Truran} J.~W., 2005, ApJ, 618, 68

\bibitem[{{Lauroesch} {et~al}\mbox{.}(1996){Lauroesch}, {Truran}, {Welty}, \&
  {York}}]{1996PASP..108..641L}
{Lauroesch} J.~T., {Truran} J.~W., {Welty} D.~E., {York} D.~G., 1996, PASP,
  108, 641

\bibitem[{{Ledoux}, {Bergeron} \& {Petitjean}(2002){Ledoux}, {Bergeron}, \&
  {Petitjean}}]{2002A&A...385..802L}
{Ledoux} C., {Bergeron} J., {Petitjean} P., 2002, A\&A, 385, 802

\bibitem[{{Ledoux}, {Petitjean} \& {Srianand}(2003){Ledoux}, {Petitjean}, \&
  {Srianand}}]{2003MNRAS.346..209L}
{Ledoux} C., {Petitjean} P., {Srianand} R., 2003, \mnras, 346, 209

\bibitem[{{Ledoux}, {Srianand} \& {Petitjean}(2002){Ledoux}, {Srianand}, \&
  {Petitjean}}]{2002A&A...392..781L}
{Ledoux} C., {Srianand} R., {Petitjean} P., 2002, \aap, 392, 781

\bibitem[{{Lee-Brown} {et~al}\mbox{.}(2015){Lee-Brown}, {Anthony-Twarog},
  {Deliyannis}, {Rich}, \& {Twarog}}]{2015AJ....149..121L}
{Lee-Brown} D.~B., {Anthony-Twarog} B.~J., {Deliyannis} C.~P., {Rich} E.,
  {Twarog} B.~A., 2015, AJ, 149, 121

\bibitem[{{Lehner} {et~al}\mbox{.}(2014){Lehner}, {O'Meara}, {Fox}, {Howk},
  {Prochaska}, {Burns}, \& {Armstrong}}]{2014ApJ...788..119L}
{Lehner} N., {O'Meara} J.~M., {Fox} A.~J., {Howk} J.~C., {Prochaska} J.~X.,
  {Burns} V., {Armstrong} A.~A., 2014, \apj, 788, 119

\bibitem[{{Lu} {et~al}\mbox{.}(1996){Lu}, {Sargent}, {Barlow}, {Churchill}, \&
  {Vogt}}]{1996ApJS..107..475L}
{Lu} L., {Sargent} W.~L.~W., {Barlow} T.~A., {Churchill} C.~W., {Vogt} S.~S.,
  1996, ApJS, 107, 475

\bibitem[{{Mathis}, {Mezger} \& {Panagia}(1983){Mathis}, {Mezger}, \&
  {Panagia}}]{1983A&A...128..212M}
{Mathis} J.~S., {Mezger} P.~G., {Panagia} N., 1983, A\&A, 128, 212

\bibitem[{{Mazzotta} {et~al}\mbox{.}(1998){Mazzotta}, {Mazzitelli},
  {Colafrancesco}, \& {Vittorio}}]{1998A&AS..133..403M}
{Mazzotta} P., {Mazzitelli} G., {Colafrancesco} S., {Vittorio} N., 1998, A\&AS,
  133, 403

\bibitem[{{Murphy} \& {Liske}(2004)}]{2004MNRAS.354L..31M}
{Murphy} M.~T., {Liske} J., 2004, MNRAS, 354, L31

\bibitem[{{Nestor} {et~al}\mbox{.}(2008){Nestor}, {Pettini}, {Hewett}, {Rao},
  \& {Wild}}]{2008MNRAS.390.1670N}
{Nestor} D.~B., {Pettini} M., {Hewett} P.~C., {Rao} S., {Wild} V., 2008, MNRAS,
  390, 1670

\bibitem[{{Nestor} {et~al}\mbox{.}(2003){Nestor}, {Rao}, {Turnshek}, \& {Vanden
  Berk}}]{2003ApJ...595L...5N}
{Nestor} D.~B., {Rao} S.~M., {Turnshek} D.~A., {Vanden Berk} D., 2003, ApJL,
  595, L5

\bibitem[{{Nestor}, {Turnshek} \& {Rao}(2005){Nestor}, {Turnshek}, \&
  {Rao}}]{2005ApJ...628..637N}
{Nestor} D.~B., {Turnshek} D.~A., {Rao} S.~M., 2005, ApJ, 628, 637

\bibitem[{{Noterdaeme} {et~al}\mbox{.}(2008){Noterdaeme}, {Ledoux},
  {Petitjean}, \& {Srianand}}]{2008A&A...481..327N}
{Noterdaeme} P., {Ledoux} C., {Petitjean} P., {Srianand} R., 2008, \aap, 481,
  327

\bibitem[{{Noterdaeme} {et~al}\mbox{.}(2010){Noterdaeme}, {Petitjean},
  {Ledoux}, {L{\'o}pez}, {Srianand}, \& {Vergani}}]{2010A&A...523A..80N}
{Noterdaeme} P., {Petitjean} P., {Ledoux} C., {L{\'o}pez} S., {Srianand} R.,
  {Vergani} S.~D., 2010, \aap, 523, A80

\bibitem[{{P{\^a}ris} {et~al}\mbox{.}(2012){P{\^a}ris}, {Petitjean}, {Aubourg},
  {Bailey}, {Ross}, {Myers}, {Strauss}, {Anderson}, {Arnau}, {Bautista},
  {Bizyaev}, {Bolton}, {Bovy}, {Brandt}, {Brewington}, {Browstein}, {Busca},
  {Capellupo}, {Carithers}, {Croft}, {Dawson}, {Delubac}, {Ebelke},
  {Eisenstein}, {Engelke}, {Fan}, {Filiz Ak}, {Finley}, {Font-Ribera}, {Ge},
  {Gibson}, {Hall}, {Hamann}, {Hennawi}, {Ho}, {Hogg}, {Ivezi{\'c}}, {Jiang},
  {Kimball}, {Kirkby}, {Kirkpatrick}, {Lee}, {Le Goff}, {Lundgren}, {MacLeod},
  {Malanushenko}, {Malanushenko}, {Maraston}, {McGreer}, {McMahon},
  {Miralda-Escud{\'e}}, {Muna}, {Noterdaeme}, {Oravetz},
  {Palanque-Delabrouille}, {Pan}, {Perez-Fournon}, {Pieri}, {Richards},
  {Rollinde}, {Sheldon}, {Schlegel}, {Schneider}, {Slosar}, {Shelden}, {Shen},
  {Simmons}, {Snedden}, {Suzuki}, {Tinker}, {Viel}, {Weaver}, {Weinberg},
  {White}, {Wood-Vasey}, \& {Y{\`e}che}}]{2012A&A...548A..66P}
{P{\^a}ris} I. {et~al.}, 2012, A\&A, 548, A66

\bibitem[{{Petitjean} {et~al}\mbox{.}(2006){Petitjean}, {Ledoux}, {Noterdaeme},
  \& {Srianand}}]{2006A&A...456L...9P}
{Petitjean} P., {Ledoux} C., {Noterdaeme} P., {Srianand} R., 2006, \aap, 456,
  L9

\bibitem[{{Petitjean}, {Srianand} \& {Ledoux}(2000){Petitjean}, {Srianand}, \&
  {Ledoux}}]{2000A&A...364L..26P}
{Petitjean} P., {Srianand} R., {Ledoux} C., 2000, \aap, 364, L26

\bibitem[{{Pettini} {et~al}\mbox{.}(1999){Pettini}, {Ellison}, {Steidel}, \&
  {Bowen}}]{1999ApJ...510..576P}
{Pettini} M., {Ellison} S.~L., {Steidel} C.~C., {Bowen} D.~V., 1999, ApJ, 510,
  576

\bibitem[{{Pettini} {et~al}\mbox{.}(1997){Pettini}, {Smith}, {King}, \&
  {Hunstead}}]{1997ApJ...486..665P}
{Pettini} M., {Smith} L.~J., {King} D.~L., {Hunstead} R.~W., 1997, ApJ, 486,
  665

\bibitem[{{Pieri} {et~al}\mbox{.}(2010){Pieri}, {Frank}, {Weinberg}, {Mathur},
  \& {York}}]{2010ApJ...724L..69P}
{Pieri} M.~M., {Frank} S., {Weinberg} D.~H., {Mathur} S., {York} D.~G., 2010,
  ApJL, 724, L69

\bibitem[{{Prochaska}, {Chen} \& {Bloom}(2006){Prochaska}, {Chen}, \&
  {Bloom}}]{2006ApJ...648...95P}
{Prochaska} J.~X., {Chen} H.-W., {Bloom} J.~S., 2006, ApJ, 648, 95

\bibitem[{{Prochaska} {et~al}\mbox{.}(2011){Prochaska}, {Weiner}, {Chen},
  {Mulchaey}, \& {Cooksey}}]{2011ApJ...740...91P}
{Prochaska} J.~X., {Weiner} B., {Chen} H.-W., {Mulchaey} J., {Cooksey} K.,
  2011, \apj, 740, 91

\bibitem[{{Prochaska} \& {Wolfe}(2002)}]{2002ApJ...566...68P}
{Prochaska} J.~X., {Wolfe} A.~M., 2002, ApJ, 566, 68

\bibitem[{{Quider} {et~al}\mbox{.}(2011){Quider}, {Nestor}, {Turnshek}, {Rao},
  {Monier}, {Weyant}, \& {Busche}}]{2011AJ....141..137Q}
{Quider} A.~M., {Nestor} D.~B., {Turnshek} D.~A., {Rao} S.~M., {Monier} E.~M.,
  {Weyant} A.~N., {Busche} J.~R., 2011, AJ, 141, 137

\bibitem[{{Richter} {et~al}\mbox{.}(2011){Richter}, {Krause}, {Fechner},
  {Charlton}, \& {Murphy}}]{2011A&A...528A..12R}
{Richter} P., {Krause} F., {Fechner} C., {Charlton} J.~C., {Murphy} M.~T.,
  2011, A\&A, 528, A12

\bibitem[{{Rousseeuw} \& {Croux}(1993)}]{1993JASA...88..1273P}
{Rousseeuw} J.~P., {Croux} C., 1993, JASA, 88, 1273

\bibitem[{{Routly} \& {Spitzer}(1952)}]{1952ApJ...115..227R}
{Routly} P.~M., {Spitzer}, Jr. L., 1952, ApJ, 115, 227

\bibitem[{{Sardane}, {Turnshek} \& {Rao}(2014){Sardane}, {Turnshek}, \&
  {Rao}}]{2014MNRAS.444.1747S}
{Sardane} G.~M., {Turnshek} D.~A., {Rao} S.~M., 2014, MNRAS, 444, 1747 Paper I

\bibitem[{{Savage} {et~al}\mbox{.}(2014){Savage}, {Kim}, {Wakker}, {Keeney},
  {Shull}, {Stocke}, \& {Green}}]{2014ApJS..212....8S}
{Savage} B.~D., {Kim} T.-S., {Wakker} B.~P., {Keeney} B., {Shull} J.~M.,
  {Stocke} J.~T., {Green} J.~C., 2014, \apjs, 212, 8

\bibitem[{{Savage} \& {Sembach}(1996)}]{1996ARA&A..34..279S}
{Savage} B.~D., {Sembach} K.~R., 1996, ARA\&A, 34, 279

\bibitem[{{Schlegel} {et~al}\mbox{.}(2007){Schlegel}, {Blanton}, {Eisenstein},
  {Gillespie}, {Gunn}, {Harding}, {McDonald}, {Nichol}, {Padmanabhan},
  {Percival}, {Richards}, {Rockosi}, {Roe}, {Ross}, {Schneider}, {Strauss},
  {Weinberg}, \& {White}}]{2007AAS...21113229S}
{Schlegel} D.~J. {et~al.}, 2007, in Bulletin of the American Astronomical
  Society, Vol.~39, American Astronomical Society Meeting Abstracts, p. 132.29

\bibitem[{{Schneider} {et~al}\mbox{.}(2010){Schneider}, {Richards}, {Hall},
  {Strauss}, {Anderson}, {Boroson}, {Ross}, {Shen}, {Brandt}, {Fan}, {Inada},
  {Jester}, {Knapp}, {Krawczyk}, {Thakar}, {Vanden Berk}, {Voges}, {Yanny},
  {York}, {Bahcall}, {Bizyaev}, {Blanton}, {Brewington}, {Brinkmann},
  {Eisenstein}, {Frieman}, {Fukugita}, {Gray}, {Gunn}, {Hibon}, {Ivezi{\'c}},
  {Kent}, {Kron}, {Lee}, {Lupton}, {Malanushenko}, {Malanushenko}, {Oravetz},
  {Pan}, {Pier}, {Price}, {Saxe}, {Schlegel}, {Simmons}, {Snedden}, {SubbaRao},
  {Szalay}, \& {Weinberg}}]{2010AJ....139.2360S}
{Schneider} D.~P. {et~al.}, 2010, AJ, 139, 2360

\bibitem[{{Seyffert} {et~al}\mbox{.}(2013){Seyffert}, {Cooksey}, {Simcoe},
  {O'Meara}, {Kao}, \& {Prochaska}}]{2013ApJ...779..161S}
{Seyffert} E.~N., {Cooksey} K.~L., {Simcoe} R.~A., {O'Meara} J.~M., {Kao}
  M.~M., {Prochaska} J.~X., 2013, ApJ, 779, 161

\bibitem[{{Shen} {et~al}\mbox{.}(2011){Shen}, {Richards}, {Strauss}, {Hall},
  {Schneider}, {Snedden}, {Bizyaev}, {Brewington}, {Malanushenko},
  {Malanushenko}, {Oravetz}, {Pan}, \& {Simmons}}]{2011ApJS..194...45S}
{Shen} Y. {et~al.}, 2011, ApJS, 194, 45

\bibitem[{{Siluk} \& {Silk}(1974)}]{1974ApJ...192...51S}
{Siluk} R.~S., {Silk} J., 1974, ApJ, 192, 51

\bibitem[{{Smee} {et~al}\mbox{.}(2013){Smee}, {Gunn}, {Uomoto}, {Roe},
  {Schlegel}, {Rockosi}, {Carr}, {Leger}, {Dawson}, {Olmstead}, {Brinkmann},
  {Owen}, {Barkhouser}, {Honscheid}, {Harding}, {Long}, {Lupton}, {Loomis},
  {Anderson}, {Annis}, {Bernardi}, {Bhardwaj}, {Bizyaev}, {Bolton},
  {Brewington}, {Briggs}, {Burles}, {Burns}, {Castander}, {Connolly},
  {Davenport}, {Ebelke}, {Epps}, {Feldman}, {Friedman}, {Frieman}, {Heckman},
  {Hull}, {Knapp}, {Lawrence}, {Loveday}, {Mannery}, {Malanushenko},
  {Malanushenko}, {Merrelli}, {Muna}, {Newman}, {Nichol}, {Oravetz}, {Pan},
  {Pope}, {Ricketts}, {Shelden}, {Sandford}, {Siegmund}, {Simmons}, {Smith},
  {Snedden}, {Schneider}, {SubbaRao}, {Tremonti}, {Waddell}, \&
  {York}}]{2013AJ....146...32S}
{Smee} S.~A. {et~al.}, 2013, AJ, 146, 32

\bibitem[{{Srianand} {et~al}\mbox{.}(2005){Srianand}, {Petitjean}, {Ledoux},
  {Ferland}, \& {Shaw}}]{2005MNRAS.362..549S}
{Srianand} R., {Petitjean} P., {Ledoux} C., {Ferland} G., {Shaw} G., 2005,
  \mnras, 362, 549

\bibitem[{{Stocke} {et~al}\mbox{.}(2014){Stocke}, {Keeney}, {Danforth},
  {Syphers}, {Yamamoto}, {Shull}, {Green}, {Froning}, {Savage}, {Wakker},
  {Kim}, {Ryan-Weber}, \& {Kacprzak}}]{2014ApJ...791..128S}
{Stocke} J.~T. {et~al.}, 2014, \apj, 791, 128

\bibitem[{{Thom} \& {Chen}(2008)}]{2008ApJ...683...22T}
{Thom} C., {Chen} H.-W., 2008, \apj, 683, 22

\bibitem[{{Tripp} {et~al}\mbox{.}(2008){Tripp}, {Sembach}, {Bowen}, {Savage},
  {Jenkins}, {Lehner}, \& {Richter}}]{2008ApJS..177...39T}
{Tripp} T.~M., {Sembach} K.~R., {Bowen} D.~V., {Savage} B.~D., {Jenkins} E.~B.,
  {Lehner} N., {Richter} P., 2008, \apjs, 177, 39

\bibitem[{{Tumlinson} {et~al}\mbox{.}(2011){Tumlinson}, {Thom}, {Werk},
  {Prochaska}, {Tripp}, {Weinberg}, {Peeples}, {O'Meara}, {Oppenheimer},
  {Meiring}, {Katz}, {Dav{\'e}}, {Ford}, \& {Sembach}}]{2011Sci...334..948T}
{Tumlinson} J. {et~al.}, 2011, Science, 334, 948

\bibitem[{{Turnshek} {et~al}\mbox{.}(1989){Turnshek}, {Wolfe}, {Lanzetta},
  {Briggs}, {Cohen}, {Foltz}, {Smith}, \& {Wilkes}}]{1989ApJ...344..567T}
{Turnshek} D.~A., {Wolfe} A.~M., {Lanzetta} K.~M., {Briggs} F.~H., {Cohen}
  R.~D., {Foltz} C.~B., {Smith} H.~E., {Wilkes} B.~J., 1989, ApJ, 344, 567

\bibitem[{{Vanden Berk} {et~al}\mbox{.}(2001){Vanden Berk}, {Richards},
  {Bauer}, {Strauss}, {Schneider}, {Heckman}, {York}, {Hall}, {Fan}, {Knapp},
  {Anderson}, {Annis}, {Bahcall}, {Bernardi}, {Briggs}, {Brinkmann}, {Brunner},
  {Burles}, {Carey}, {Castander}, {Connolly}, {Crocker}, {Csabai}, {Doi},
  {Finkbeiner}, {Friedman}, {Frieman}, {Fukugita}, {Gunn}, {Hennessy},
  {Ivezi{\'c}}, {Kent}, {Kunszt}, {Lamb}, {Leger}, {Long}, {Loveday}, {Lupton},
  {Meiksin}, {Merelli}, {Munn}, {Newberg}, {Newcomb}, {Nichol}, {Owen}, {Pier},
  {Pope}, {Rockosi}, {Schlegel}, {Siegmund}, {Smee}, {Snir}, {Stoughton},
  {Stubbs}, {SubbaRao}, {Szalay}, {Szokoly}, {Tremonti}, {Uomoto}, {Waddell},
  {Yanny}, \& {Zheng}}]{2001AJ....122..549V}
{Vanden Berk} D.~E. {et~al.}, 2001, AJ, 122, 549

\bibitem[{{Verner}(1999)}]{1999PhST...83..174V}
{Verner} D.~A., 1999, PhST, 83, 174

\bibitem[{{Verner} {et~al}\mbox{.}(1996){Verner}, {Ferland}, {Korista}, \&
  {Yakovlev}}]{1996ApJ...465..487V}
{Verner} D.~A., {Ferland} G.~J., {Korista} K.~T., {Yakovlev} D.~G., 1996, ApJ,
  465, 487

\bibitem[{{Viegas}(1995)}]{1995MNRAS.276..268V}
{Viegas} S.~M., 1995, MNRAS, 276, 268

\bibitem[{{Vladilo} {et~al}\mbox{.}(2001){Vladilo}, {Centuri{\'o}n},
  {Bonifacio}, \& {Howk}}]{2001ApJ...557.1007V}
{Vladilo} G., {Centuri{\'o}n} M., {Bonifacio} P., {Howk} J.~C., 2001, ApJ, 557,
  1007

\bibitem[{{Vladilo}, {Prochaska} \& {Wolfe}(2008){Vladilo}, {Prochaska}, \&
  {Wolfe}}]{2008A&A...478..701V}
{Vladilo} G., {Prochaska} J.~X., {Wolfe} A.~M., 2008, A\&A, 478, 701

\bibitem[{{Welty} \& {Crowther}(2010)}]{2010MNRAS.404.1321W}
{Welty} D.~E., {Crowther} P.~A., 2010, MNRAS, 404, 1321

\bibitem[{{Welty} {et~al}\mbox{.}(1999){Welty}, {Hobbs}, {Lauroesch}, {Morton},
  {Spitzer}, \& {York}}]{1999ApJS..124..465W}
{Welty} D.~E., {Hobbs} L.~M., {Lauroesch} J.~T., {Morton} D.~C., {Spitzer} L.,
  {York} D.~G., 1999, ApJS, 124, 465

\bibitem[{{Welty} {et~al}\mbox{.}(2001){Welty}, {Lauroesch}, {Blades}, {Hobbs},
  \& {York}}]{2001ApJ...554L..75W}
{Welty} D.~E., {Lauroesch} J.~T., {Blades} J.~C., {Hobbs} L.~M., {York} D.~G.,
  2001, ApJL, 554, L75

\bibitem[{{Welty}, {Morton} \& {Hobbs}(1996){Welty}, {Morton}, \&
  {Hobbs}}]{1996ApJS..106..533W}
{Welty} D.~E., {Morton} D.~C., {Hobbs} L.~M., 1996, ApJS, 106, 533

\bibitem[{{Wild}, {Hewett} \& {Pettini}(2006){Wild}, {Hewett}, \&
  {Pettini}}]{2006MNRAS.367..211W}
{Wild} V., {Hewett} P.~C., {Pettini} M., 2006, MNRAS, 367, 211

\bibitem[{{Wolfe} \& {Prochaska}(2000)}]{2000ApJ...545..591W}
{Wolfe} A.~M., {Prochaska} J.~X., 2000, \apj, 545, 591

\bibitem[{{York} {et~al}\mbox{.}(2006){York}, {Khare}, {Vanden Berk},
  {Kulkarni}, {Crotts}, {Lauroesch}, {Richards}, {Schneider}, {Welty},
  {Alsayyad}, {Kumar}, {Lundgren}, {Shanidze}, {Smith}, {Vanlandingham},
  {Baugher}, {Hall}, {Jenkins}, {Menard}, {Rao}, {Tumlinson}, {Turnshek},
  {Yip}, \& {Brinkmann}}]{2006MNRAS.367..945Y}
{York} D.~G. {et~al.}, 2006, MNRAS, 367, 945

\bibitem[{{Zhu} \& {M{\'e}nard}(2013)}]{2013ApJ...770..130Z}
{Zhu} G., {M{\'e}nard} B., 2013, ApJ, 770, 130

\bibitem[{{Zych} {et~al}\mbox{.}(2009){Zych}, {Murphy}, {Hewett}, \&
  {Prochaska}}]{2009MNRAS.392.1429Z}
{Zych} B.~J., {Murphy} M.~T., {Hewett} P.~C., {Prochaska} J.~X., 2009, MNRAS,
  392, 1429

\end{thebibliography}
